\documentstyle[epic,eepic,aps,epsfig,pre,twocolumn]{revtex}
  \title{Boundary Effects on the Angle of Repose in Rotating Cylinders}
  \author{Christian M. Dury and Gerald H. Ristow}
  \address{Fachbereich Physik, Philipps-Universit\"at, Renthof 6, 
           35032 Marburg, Germany}
  \author{Jamie L. Moss and Masami Nakagawa}
  \address{Division of Engineering, Colorado School of Mines, Golden, CO 80401, 
           USA}
  \date{received 29 August 1997; revised \today}
  \draft
\begin{document}
\maketitle
\begin{abstract}
The angle of repose for the flow of granular materials in a half-filled
rotating drum is studied by means of experiments and computer simulations.
Particles of different material properties are used to investigate the effects
of the end caps on the angle of repose. By fitting the numerical results to an
exponentially decaying function, we are able to calculate the characteristic 
range, $\zeta$, of the influence of the wall. We found that $\zeta$ scales
with  the drum radius but does not depend on either the density or the
gravitational constant. For increasing particle diameter, finite size effects
are visible.
\end{abstract}

\section{Introduction}
The behavior of granular materials is of great technological interest 
\cite{jaeger96} and its investigation has a history of more than two hundred 
years. When granular materials are put in a rotating drum, avalanches are 
observed along the surface of the granular bulk~\cite{evesque88,rajchenbach90}.
In industrial processes, such devices are mostly used for mixing different
kinds of particles. However, it is also well known that particles of different
sizes tend to segregate in the radial and axial directions~\cite{donald62,%
bridgwater76,dasgupta91,nakagawa94,clement95,cantelaube95,hill94,hill95,%
hill97,nakagawa97b,hill97b}.

Recently, the particle dynamics of granular materials in a rotating drum has
been described by using quasi two-dimensional systems, tracking  individual
grains via cameras and computer programs~\cite{cantelaube95}.  Extensive
numerical studies have also reproduced and predicted many of the experimental
findings~\cite{walton93,ristow94,baumann94,baumann95,buchholtz95,ristow96,%
dury97}. The segregation and mixing process depends on many parameters, such 
as size~\cite{bridgwater76,dury97}, shape~\cite{buchholtz95},
mass~\cite{ristow94}, frictional forces, angular velocity~\cite{dury97},
filling level of the drum~\cite{metcalfe95}, etc. The angle of repose of the
material also depends on the parameters and it was argued that either the
dynamic or static angle difference of the materials in the drum influence the
axial segregation process~\cite{donald62,dasgupta91,hill94,hill95}.

In this article, we investigate experimentally the dependence of the dynamic
angle of repose on the rotation speed of a half-filled drum for particles of
different material properties. It is found that the angle is up to 5 degrees
higher at the end caps of the drum due to boundary friction. Using a
three-dimensional discrete element code, we are able to quantify this boundary
effect and discuss its dependence on gravity, particle size and density.

\section{Experimental Results}
\label{experiment}
An acrylic cylinder of diameter 6.9 cm and length 49 cm was placed
horizontally on two sets of roller supports and was rotated by a
well-regulated electronic motor. The material used was mustard seeds which are
relatively round of average diameter about 2.5 mm, and have a coefficient of
restitution, estimated from a set of impact experiments, of about
0.75~\cite{nakagawa93}. A set of experiments were conducted to measure the
angle of repose in different flow regimes. For a small rotation speed,
$\Omega$, intermittent flow led to a different angle before and after each
avalanche occurred, called the starting (maximum) and stopping (minimum)
angle, respectively. For a larger rotation speed these intermittent 
avalanches became a continuous flat surface and thus enabled to define one
angle of repose defined as the {\em dynamic angle of repose} as shown in
Fig.~\ref{fig: sketch}a. When $\Omega$ increases, the flat surface deforms with
increasing rotation speeds and develops a so-called S-shape surface for 
higher rotation speeds, shown in Fig.~\ref{fig: sketch}c. The 
deformation mostly starts from the lower boundary inwards and can be well 
approximated by two straight lines with different slopes close to this 
transition, sketched in Fig.~\ref{fig: sketch}b. For all measurements in this 
regime, we took the slope of the line to the right which corresponds to
the line with the higher slope.
\ifx\grdraft\undefined
\else
\vspace{-2ex}
\begin{figure}[htb]
  \begin{center}
    \epsfig{file=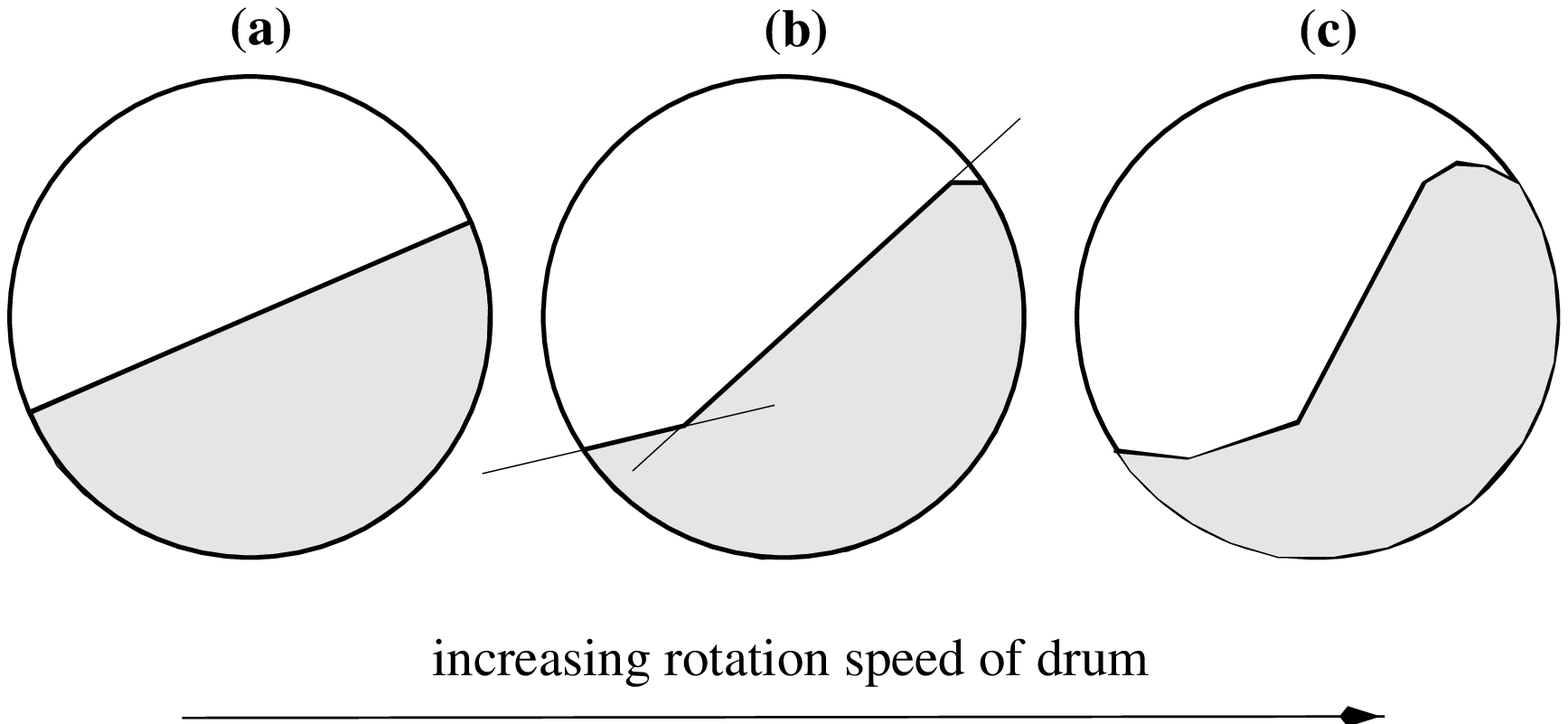,width=0.5\textwidth}
  \end{center}
  \caption{(a) Flat surface for low rotation speeds, (b) deformed surface for
           medium rotation speeds with two straight lines added as approximation
	   and (c) fully developed S-shaped surface for higher rotation speeds.}
  \label{fig: sketch}
\end{figure}
\fi

The average maximum and minimum angles of repose for the intermittent
avalanches were found to be about 36 and 32 degrees, respectively, see
Fig.~\ref{fig: mustard_all}. There seems to be a rather sharp transition from
intermittent to continuous avalanches, which happens around $\Omega = 4$ rpm.
For $\Omega$ greater than 4 rpm where the avalanches are continuous, the 
mustard seed data indicate a linear dependence of the dynamic angle of repose
on the rotation speed which differs from the quadratic dependence found by
Rajchenbach~\cite{rajchenbach90}.
\ifx\grdraft\undefined
\else
\vspace{-4ex}
\begin{figure}[htb]
  \begin{center}
\setlength{\unitlength}{0.240900pt}
\begin{picture}(1049,809)(0,0)
\thicklines \path(220,113)(240,113)
\thicklines \path(985,113)(965,113)
\put(198,113){\makebox(0,0)[r]{26}}
\thicklines \path(220,281)(240,281)
\thicklines \path(985,281)(965,281)
\put(198,281){\makebox(0,0)[r]{30}}
\thicklines \path(220,450)(240,450)
\thicklines \path(985,450)(965,450)
\put(198,450){\makebox(0,0)[r]{34}}
\thicklines \path(220,618)(240,618)
\thicklines \path(985,618)(965,618)
\put(198,618){\makebox(0,0)[r]{38}}
\thicklines \path(220,786)(240,786)
\thicklines \path(985,786)(965,786)
\put(198,786){\makebox(0,0)[r]{42}}
\thicklines \path(220,113)(220,133)
\thicklines \path(220,786)(220,766)
\put(220,68){\makebox(0,0){0}}
\thicklines \path(411,113)(411,133)
\thicklines \path(411,786)(411,766)
\put(411,68){\makebox(0,0){5}}
\thicklines \path(603,113)(603,133)
\thicklines \path(603,786)(603,766)
\put(603,68){\makebox(0,0){10}}
\thicklines \path(794,113)(794,133)
\thicklines \path(794,786)(794,766)
\put(794,68){\makebox(0,0){15}}
\thicklines \path(985,113)(985,133)
\thicklines \path(985,786)(985,766)
\put(985,68){\makebox(0,0){20}}
\thicklines \path(220,113)(985,113)(985,786)(220,786)(220,113)
\put(45,449){\makebox(0,0)[l]{\shortstack{$\langle\Theta\rangle$}}}
\put(602,0){\makebox(0,0){$\Omega$ [rpm]}}
\put(243,481){\circle{18}}
\put(252,505){\circle{18}}
\put(263,534){\circle{18}}
\put(266,500){\circle{18}}
\put(277,555){\circle{18}}
\put(284,492){\circle{18}}
\put(293,502){\circle{18}}
\put(307,475){\circle{18}}
\put(319,502){\circle{18}}
\put(333,492){\circle{18}}
\put(346,550){\circle{18}}
\thinlines \path(243,446)(243,516)
\thinlines \path(233,446)(253,446)
\thinlines \path(233,516)(253,516)
\thinlines \path(252,486)(252,525)
\thinlines \path(242,486)(262,486)
\thinlines \path(242,525)(262,525)
\thinlines \path(263,534)(263,534)
\thinlines \path(253,534)(273,534)
\thinlines \path(253,534)(273,534)
\thinlines \path(266,468)(266,532)
\thinlines \path(256,468)(276,468)
\thinlines \path(256,532)(276,532)
\thinlines \path(277,534)(277,576)
\thinlines \path(267,534)(287,534)
\thinlines \path(267,576)(287,576)
\thinlines \path(284,492)(284,492)
\thinlines \path(274,492)(294,492)
\thinlines \path(274,492)(294,492)
\thinlines \path(293,484)(293,520)
\thinlines \path(283,484)(303,484)
\thinlines \path(283,520)(303,520)
\thinlines \path(307,454)(307,495)
\thinlines \path(297,454)(317,454)
\thinlines \path(297,495)(317,495)
\thinlines \path(319,484)(319,520)
\thinlines \path(309,484)(329,484)
\thinlines \path(309,520)(329,520)
\thinlines \path(333,492)(333,492)
\thinlines \path(323,492)(343,492)
\thinlines \path(323,492)(343,492)
\thinlines \path(346,530)(346,571)
\thinlines \path(336,530)(356,530)
\thinlines \path(336,571)(356,571)
\put(243,239){\makebox(0,0){$\star$}}
\put(252,379){\makebox(0,0){$\star$}}
\put(263,323){\makebox(0,0){$\star$}}
\put(266,292){\makebox(0,0){$\star$}}
\put(277,334){\makebox(0,0){$\star$}}
\put(284,309){\makebox(0,0){$\star$}}
\put(293,334){\makebox(0,0){$\star$}}
\put(307,340){\makebox(0,0){$\star$}}
\put(319,334){\makebox(0,0){$\star$}}
\put(333,323){\makebox(0,0){$\star$}}
\put(346,365){\makebox(0,0){$\star$}}
\thinlines \path(243,209)(243,269)
\thinlines \path(233,209)(253,209)
\thinlines \path(233,269)(253,269)
\thinlines \path(252,359)(252,399)
\thinlines \path(242,359)(262,359)
\thinlines \path(242,399)(262,399)
\thinlines \path(263,323)(263,323)
\thinlines \path(253,323)(273,323)
\thinlines \path(253,323)(273,323)
\thinlines \path(266,274)(266,310)
\thinlines \path(256,274)(276,274)
\thinlines \path(256,310)(276,310)
\thinlines \path(277,299)(277,369)
\thinlines \path(267,299)(287,299)
\thinlines \path(267,369)(287,369)
\thinlines \path(284,290)(284,329)
\thinlines \path(274,290)(294,290)
\thinlines \path(274,329)(294,329)
\thinlines \path(293,299)(293,369)
\thinlines \path(283,299)(303,299)
\thinlines \path(283,369)(303,369)
\thinlines \path(307,320)(307,361)
\thinlines \path(297,320)(317,320)
\thinlines \path(297,361)(317,361)
\thinlines \path(319,316)(319,352)
\thinlines \path(309,316)(329,316)
\thinlines \path(309,352)(329,352)
\thinlines \path(333,323)(333,323)
\thinlines \path(323,323)(343,323)
\thinlines \path(323,323)(343,323)
\thinlines \path(346,339)(346,392)
\thinlines \path(336,339)(356,339)
\thinlines \path(336,392)(356,392)
\put(349,460){\circle*{12}}
\put(356,466){\circle*{12}}
\put(365,475){\circle*{12}}
\put(367,450){\circle*{12}}
\put(415,492){\circle*{12}}
\put(495,534){\circle*{12}}
\put(633,576){\circle*{12}}
\put(748,618){\circle*{12}}
\put(909,660){\circle*{12}}
\thinlines \path(349,442)(349,478)
\thinlines \path(339,442)(359,442)
\thinlines \path(339,478)(359,478)
\thinlines \path(356,433)(356,500)
\thinlines \path(346,433)(366,433)
\thinlines \path(346,500)(366,500)
\thinlines \path(365,454)(365,495)
\thinlines \path(355,454)(375,454)
\thinlines \path(355,495)(375,495)
\thinlines \path(367,420)(367,479)
\thinlines \path(357,420)(377,420)
\thinlines \path(357,479)(377,479)
\thinlines \path(415,492)(415,492)
\thinlines \path(405,492)(425,492)
\thinlines \path(405,492)(425,492)
\thinlines \path(495,534)(495,534)
\thinlines \path(485,534)(505,534)
\thinlines \path(485,534)(505,534)
\thinlines \path(633,576)(633,576)
\thinlines \path(623,576)(643,576)
\thinlines \path(623,576)(643,576)
\thinlines \path(748,618)(748,618)
\thinlines \path(738,618)(758,618)
\thinlines \path(738,618)(758,618)
\thinlines \path(909,660)(909,660)
\thinlines \path(899,660)(919,660)
\thinlines \path(899,660)(919,660)
\thicklines \path(349,460)(349,460)(356,466)(365,475)(367,450)(415,492)(495,534)(633,576)(748,618)(909,660)
\end{picture}
  \end{center}
  \caption{Experimentally measured starting ($\circ$) and stopping ($\star$) 
           angle and dynamic angle of repose ($\bullet$) for mustard seeds.}
  \label{fig: mustard_all}
\end{figure}
\fi

We also investigated the dynamic angle of repose for different particle
diameters and materials in the continuous regime in more detail using a 27 cm
long acrylic cylinder of diameter $2R=6.9$ cm. For a given rotation speed,
$\Omega$, the dynamic angle of repose was measured four times at one of the
acrylic end caps and the average value with an error bar corresponding to a
confidence interval of 2$\sigma$, where $\sigma$ is the standard deviation of
the data points, was then calculated. First we used mustard seeds of two
different diameters, namely 1.7 mm (black) and 2.5 mm (yellow), with a density
of 1.3 g/cm$^3$. The latter were the same that were used to produce
Fig.~\ref{fig: mustard_all}. We varied the rotation speed, $\Omega$, from 5 rpm to 40 rpm
and took the higher angle in the S-shaped regime which exists for higher
rotation rates, see Fig.~\ref{fig: sketch}b. Both data sets are shown in
Fig.~\ref{fig: mustard} for black ($\bullet$) and yellow ($\circ$) seeds. The
figure also illustrates the transition to the S-shaped regime which occurs at
the change of slope, e.g.\ at around 11 rpm for the smaller seeds and around 16
rpm for the larger seeds. One also notes that the dynamic angle of repose is
much higher for the larger particles in the low frequency regime. For values of
$\Omega > 15$ rpm in the S-shaped regime, the difference in the dynamic angle
of repose for the two different types of mustard seeds decreases with
increasing $\Omega$, and both curves cross around 30 rpm giving a slightly
higher angle for the smaller seeds with the highest rotation speeds studied.

We applied the same measurements to two sets of glass beads having a density of
2.6 g/cm$^3$. The smaller beads had a diameter of 1.5 mm with no measurable
size distribution, whereas the larger beads had a diameter range of 3.0 $\pm$
0.2 mm. Both data sets are shown in Fig.~\ref{fig: glass} for small ($\bullet$)
and large ($\circ$) beads. It can be seen from this figure that the transition 
to the S-shaped regime occurs at around 16 rpm for the smaller beads and around
24 rpm for the larger beads. In general, we found that the small particles
exhibit the S-shaped surface at lower values  of $\Omega$ than the large
particles. The angles of repose are, in general, lower for the glass beads
compared to the mustard seeds which we attribute to the fact that the mustard
seeds are not as round as the glass beads and rotations of the mustard seeds
are therefore more suppressed. The coefficient of friction is also higher for
mustard seeds.
\ifx\grdraft\undefined
\else
\vspace{-4ex}
\begin{figure}[htb]
  \begin{center}
\setlength{\unitlength}{0.240900pt}
\begin{picture}(1049,809)(0,0)
\thicklines \path(220,113)(240,113)
\thicklines \path(985,113)(965,113)
\put(198,113){\makebox(0,0)[r]{32}}
\thicklines \path(220,281)(240,281)
\thicklines \path(985,281)(965,281)
\put(198,281){\makebox(0,0)[r]{36}}
\thicklines \path(220,450)(240,450)
\thicklines \path(985,450)(965,450)
\put(198,450){\makebox(0,0)[r]{40}}
\thicklines \path(220,618)(240,618)
\thicklines \path(985,618)(965,618)
\put(198,618){\makebox(0,0)[r]{44}}
\thicklines \path(220,786)(240,786)
\thicklines \path(985,786)(965,786)
\put(198,786){\makebox(0,0)[r]{48}}
\thicklines \path(220,113)(220,133)
\thicklines \path(220,786)(220,766)
\put(220,68){\makebox(0,0){0}}
\thicklines \path(390,113)(390,133)
\thicklines \path(390,786)(390,766)
\put(390,68){\makebox(0,0){10}}
\thicklines \path(560,113)(560,133)
\thicklines \path(560,786)(560,766)
\put(560,68){\makebox(0,0){20}}
\thicklines \path(730,113)(730,133)
\thicklines \path(730,786)(730,766)
\put(730,68){\makebox(0,0){30}}
\thicklines \path(900,113)(900,133)
\thicklines \path(900,786)(900,766)
\put(900,68){\makebox(0,0){40}}
\thicklines \path(220,113)(985,113)(985,786)(220,786)(220,113)
\put(45,449){\makebox(0,0)[l]{\shortstack{$\langle\Theta\rangle$}}}
\put(602,0){\makebox(0,0){$\Omega$ [rpm]}}
\put(475,702){\makebox(0,0)[r]{2.5 mm}}
\put(519,702){\circle{18}}
\put(305,244){\circle{18}}
\put(327,251){\circle{18}}
\put(348,259){\circle{18}}
\put(370,287){\circle{18}}
\put(390,289){\circle{18}}
\put(433,301){\circle{18}}
\put(475,360){\circle{18}}
\put(518,413){\circle{18}}
\put(560,431){\circle{18}}
\put(645,503){\circle{18}}
\put(730,560){\circle{18}}
\put(900,660){\circle{18}}
\thinlines \path(497,702)(563,702)
\thinlines \path(497,712)(497,692)
\thinlines \path(563,712)(563,692)
\thinlines \path(305,234)(305,255)
\thinlines \path(295,234)(315,234)
\thinlines \path(295,255)(315,255)
\thinlines \path(327,237)(327,265)
\thinlines \path(317,237)(337,237)
\thinlines \path(317,265)(337,265)
\thinlines \path(348,248)(348,270)
\thinlines \path(338,248)(358,248)
\thinlines \path(338,270)(358,270)
\thinlines \path(370,276)(370,297)
\thinlines \path(360,276)(380,276)
\thinlines \path(360,297)(380,297)
\thinlines \path(390,279)(390,299)
\thinlines \path(380,279)(400,279)
\thinlines \path(380,299)(400,299)
\thinlines \path(433,292)(433,311)
\thinlines \path(423,292)(443,292)
\thinlines \path(423,311)(443,311)
\thinlines \path(475,347)(475,373)
\thinlines \path(465,347)(485,347)
\thinlines \path(465,373)(485,373)
\thinlines \path(518,407)(518,418)
\thinlines \path(508,407)(528,407)
\thinlines \path(508,418)(528,418)
\thinlines \path(560,422)(560,439)
\thinlines \path(550,422)(570,422)
\thinlines \path(550,439)(570,439)
\thinlines \path(645,492)(645,514)
\thinlines \path(635,492)(655,492)
\thinlines \path(635,514)(655,514)
\thinlines \path(730,537)(730,583)
\thinlines \path(720,537)(740,537)
\thinlines \path(720,583)(740,583)
\thinlines \path(900,648)(900,672)
\thinlines \path(890,648)(910,648)
\thinlines \path(890,672)(910,672)
\put(475,657){\makebox(0,0)[r]{1.7 mm}}
\put(519,657){\circle*{12}}
\put(285,166){\circle*{12}}
\put(305,160){\circle*{12}}
\put(327,176){\circle*{12}}
\put(348,166){\circle*{12}}
\put(370,187){\circle*{12}}
\put(390,192){\circle*{12}}
\put(433,262){\circle*{12}}
\put(475,322){\circle*{12}}
\put(518,357){\circle*{12}}
\put(560,420){\circle*{12}}
\put(603,441){\circle*{12}}
\put(645,502){\circle*{12}}
\put(730,565){\circle*{12}}
\put(815,660){\circle*{12}}
\put(900,704){\circle*{12}}
\thinlines \path(497,657)(563,657)
\thinlines \path(497,667)(497,647)
\thinlines \path(563,667)(563,647)
\thinlines \path(285,160)(285,171)
\thinlines \path(275,160)(295,160)
\thinlines \path(275,171)(295,171)
\thinlines \path(305,155)(305,166)
\thinlines \path(295,155)(315,155)
\thinlines \path(295,166)(315,166)
\thinlines \path(327,164)(327,188)
\thinlines \path(317,164)(337,164)
\thinlines \path(317,188)(337,188)
\thinlines \path(348,156)(348,176)
\thinlines \path(338,156)(358,156)
\thinlines \path(338,176)(358,176)
\thinlines \path(370,181)(370,192)
\thinlines \path(360,181)(380,181)
\thinlines \path(360,192)(380,192)
\thinlines \path(390,182)(390,202)
\thinlines \path(380,182)(400,182)
\thinlines \path(380,202)(400,202)
\thinlines \path(433,255)(433,269)
\thinlines \path(423,255)(443,255)
\thinlines \path(423,269)(443,269)
\thinlines \path(475,314)(475,330)
\thinlines \path(465,314)(485,314)
\thinlines \path(465,330)(485,330)
\thinlines \path(518,347)(518,367)
\thinlines \path(508,347)(528,347)
\thinlines \path(508,367)(528,367)
\thinlines \path(560,415)(560,425)
\thinlines \path(550,415)(570,415)
\thinlines \path(550,425)(570,425)
\thinlines \path(603,429)(603,453)
\thinlines \path(593,429)(613,429)
\thinlines \path(593,453)(613,453)
\thinlines \path(645,496)(645,508)
\thinlines \path(635,496)(655,496)
\thinlines \path(635,508)(655,508)
\thinlines \path(730,559)(730,571)
\thinlines \path(720,559)(740,559)
\thinlines \path(720,571)(740,571)
\thinlines \path(815,649)(815,670)
\thinlines \path(805,649)(825,649)
\thinlines \path(805,670)(825,670)
\thinlines \path(900,697)(900,711)
\thinlines \path(890,697)(910,697)
\thinlines \path(890,711)(910,711)
\end{picture}
  \end{center}
  \caption{Dynamic angle of repose for black ($\bullet$) and yellow ($\circ$) 
           mustard seeds.}
  \label{fig: mustard}
\end{figure}
\fi
\ifx\grdraft\undefined
\else
\vspace{-4ex}
\begin{figure}[htb]
  \begin{center}
\setlength{\unitlength}{0.240900pt}
\begin{picture}(1049,809)(0,0)
\thicklines \path(220,113)(240,113)
\thicklines \path(985,113)(965,113)
\put(198,113){\makebox(0,0)[r]{28}}
\thicklines \path(220,281)(240,281)
\thicklines \path(985,281)(965,281)
\put(198,281){\makebox(0,0)[r]{32}}
\thicklines \path(220,450)(240,450)
\thicklines \path(985,450)(965,450)
\put(198,450){\makebox(0,0)[r]{36}}
\thicklines \path(220,618)(240,618)
\thicklines \path(985,618)(965,618)
\put(198,618){\makebox(0,0)[r]{40}}
\thicklines \path(220,786)(240,786)
\thicklines \path(985,786)(965,786)
\put(198,786){\makebox(0,0)[r]{44}}
\thicklines \path(220,113)(220,133)
\thicklines \path(220,786)(220,766)
\put(220,68){\makebox(0,0){0}}
\thicklines \path(390,113)(390,133)
\thicklines \path(390,786)(390,766)
\put(390,68){\makebox(0,0){10}}
\thicklines \path(560,113)(560,133)
\thicklines \path(560,786)(560,766)
\put(560,68){\makebox(0,0){20}}
\thicklines \path(730,113)(730,133)
\thicklines \path(730,786)(730,766)
\put(730,68){\makebox(0,0){30}}
\thicklines \path(900,113)(900,133)
\thicklines \path(900,786)(900,766)
\put(900,68){\makebox(0,0){40}}
\thicklines \path(220,113)(985,113)(985,786)(220,786)(220,113)
\put(45,449){\makebox(0,0)[l]{\shortstack{$\langle\Theta\rangle$}}}
\put(602,0){\makebox(0,0){$\Omega$ [rpm]}}
\put(475,702){\makebox(0,0)[r]{1.5 mm}}
\put(519,702){\circle*{12}}
\put(285,171){\circle*{12}}
\put(305,181){\circle*{12}}
\put(327,196){\circle*{12}}
\put(348,209){\circle*{12}}
\put(370,227){\circle*{12}}
\put(390,237){\circle*{12}}
\put(433,257){\circle*{12}}
\put(475,269){\circle*{12}}
\put(518,315){\circle*{12}}
\put(560,357){\circle*{12}}
\put(645,444){\circle*{12}}
\put(730,518){\circle*{12}}
\put(815,618){\circle*{12}}
\put(900,691){\circle*{12}}
\thinlines \path(497,702)(563,702)
\thinlines \path(497,712)(497,692)
\thinlines \path(563,712)(563,692)
\thinlines \path(285,162)(285,180)
\thinlines \path(275,162)(295,162)
\thinlines \path(275,180)(295,180)
\thinlines \path(305,173)(305,190)
\thinlines \path(295,173)(315,173)
\thinlines \path(295,190)(315,190)
\thinlines \path(327,192)(327,200)
\thinlines \path(317,192)(337,192)
\thinlines \path(317,200)(337,200)
\thinlines \path(348,199)(348,218)
\thinlines \path(338,199)(358,199)
\thinlines \path(338,218)(358,218)
\thinlines \path(370,220)(370,233)
\thinlines \path(360,220)(380,220)
\thinlines \path(360,233)(380,233)
\thinlines \path(390,228)(390,246)
\thinlines \path(380,228)(400,228)
\thinlines \path(380,246)(400,246)
\thinlines \path(433,249)(433,265)
\thinlines \path(423,249)(443,249)
\thinlines \path(423,265)(443,265)
\thinlines \path(475,262)(475,275)
\thinlines \path(465,262)(485,262)
\thinlines \path(465,275)(485,275)
\thinlines \path(518,305)(518,325)
\thinlines \path(508,305)(528,305)
\thinlines \path(508,325)(528,325)
\thinlines \path(560,351)(560,363)
\thinlines \path(550,351)(570,351)
\thinlines \path(550,363)(570,363)
\thinlines \path(645,429)(645,460)
\thinlines \path(635,429)(655,429)
\thinlines \path(635,460)(655,460)
\thinlines \path(730,509)(730,527)
\thinlines \path(720,509)(740,509)
\thinlines \path(720,527)(740,527)
\thinlines \path(815,605)(815,631)
\thinlines \path(805,605)(825,605)
\thinlines \path(805,631)(825,631)
\thinlines \path(900,686)(900,697)
\thinlines \path(890,686)(910,686)
\thinlines \path(890,697)(910,697)
\put(475,657){\makebox(0,0)[r]{3.0 mm}}
\put(519,657){\circle{18}}
\put(305,187){\circle{18}}
\put(327,192){\circle{18}}
\put(348,199){\circle{18}}
\put(370,222){\circle{18}}
\put(390,244){\circle{18}}
\put(433,250){\circle{18}}
\put(475,289){\circle{18}}
\put(518,297){\circle{18}}
\put(560,310){\circle{18}}
\put(603,318){\circle{18}}
\put(645,355){\circle{18}}
\put(730,413){\circle{18}}
\put(815,492){\circle{18}}
\put(900,544){\circle{18}}
\thinlines \path(497,657)(563,657)
\thinlines \path(497,667)(497,647)
\thinlines \path(563,667)(563,647)
\thinlines \path(305,173)(305,200)
\thinlines \path(295,173)(315,173)
\thinlines \path(295,200)(315,200)
\thinlines \path(327,179)(327,205)
\thinlines \path(317,179)(337,179)
\thinlines \path(317,205)(337,205)
\thinlines \path(348,190)(348,208)
\thinlines \path(338,190)(358,190)
\thinlines \path(338,208)(358,208)
\thinlines \path(370,209)(370,235)
\thinlines \path(360,209)(380,209)
\thinlines \path(360,235)(380,235)
\thinlines \path(390,239)(390,250)
\thinlines \path(380,239)(400,239)
\thinlines \path(380,250)(400,250)
\thinlines \path(433,230)(433,270)
\thinlines \path(423,230)(443,230)
\thinlines \path(423,270)(443,270)
\thinlines \path(475,277)(475,300)
\thinlines \path(465,277)(485,277)
\thinlines \path(465,300)(485,300)
\thinlines \path(518,287)(518,307)
\thinlines \path(508,287)(528,287)
\thinlines \path(508,307)(528,307)
\thinlines \path(560,305)(560,315)
\thinlines \path(550,305)(570,305)
\thinlines \path(550,315)(570,315)
\thinlines \path(603,308)(603,328)
\thinlines \path(593,308)(613,308)
\thinlines \path(593,328)(613,328)
\thinlines \path(645,327)(645,383)
\thinlines \path(635,327)(655,327)
\thinlines \path(635,383)(655,383)
\thinlines \path(730,407)(730,418)
\thinlines \path(720,407)(740,407)
\thinlines \path(720,418)(740,418)
\thinlines \path(815,481)(815,502)
\thinlines \path(805,481)(825,481)
\thinlines \path(805,502)(825,502)
\thinlines \path(900,534)(900,555)
\thinlines \path(890,534)(910,534)
\thinlines \path(890,555)(910,555)
\end{picture}
  \end{center}
  \caption{Dynamic angle of repose for small ($\bullet$) and large ($\circ$) 
           glass beads.}
  \label{fig: glass}
\end{figure}
\fi

There are two striking differences when comparing Figs.~\ref{fig: glass} (glass
spheres) and \ref{fig: mustard} (mustard seeds). For rotation speeds,
$\Omega$,  lower than 15 rpm, the small and large glass beads have the same
dynamic angle of repose which agrees with the findings in~\cite{zik94}, whereas
the dynamic angle of repose is significantly higher (3 to 4 degrees) for the
larger mustard seeds compared to the smaller ones. For rotation speeds,
$\Omega$, higher than 15 rpm, the smaller glass beads show a higher dynamic
angle of repose than the larger glass beads, and this angle difference
increases with increasing rotation speed. For mustard seeds, the difference in
the dynamic angle of repose between the smaller and the larger particles
decreases with increasing $\Omega$ and the  smaller seeds only show a higher
angle for the highest rotation speeds studied. Both Fig.~\ref{fig: mustard} and
Fig.~\ref{fig: glass} seem to indicate that the increase in the dynamic angle
of repose with rotation speed, $\Omega$, in the S-shaped regime is larger for
the smaller particles.

All the above angle of repose were measured by looking through one of the
acrylic end caps. In order to study the boundary effect of these end caps on
the dynamic angle of repose, we  performed Magnetic Resonance Imaging (MRI)
measurements. This technique of studying non-invasively the flow properties of
granular materials was first used by Nakagawa et al.~\cite{nakagawa93} and is
in addition explained in more detail in ref.~\cite{nakagawa94}. We used the
large mustard seeds, which had an average diameter of 2.5 mm. The dynamic angle
of repose was measured based on the concentration data which was averaged in a
thin cross-sectional slice in the middle of the cylinder far away from the end
caps. It is shown in Fig.~\ref{fig: mri} as function of the rotation speed by
the open circles ($\circ$). We restricted the measurement to the flat surface
regime, and all data points then lie approximately on a straight line. On the
other hand, non-MRI data were measured at the end caps and are shown as stars
($\star$) in Fig.~\ref{fig: mri}. The consistently higher dynamic angle of
repose at the end cap indicates the significance of the friction between
particles and the boundary wall. We also found that the S-shape regime seems to
start earlier at the end caps due to the additional wall friction.
\ifx\grdraft\undefined
\else
\vspace{-4ex}
\begin{figure}[htb]
  \begin{center}
\setlength{\unitlength}{0.240900pt}
\begin{picture}(1049,809)(0,0)
\thicklines \path(220,113)(240,113)
\thicklines \path(985,113)(965,113)
\put(198,113){\makebox(0,0)[r]{28}}
\thicklines \path(220,337)(240,337)
\thicklines \path(985,337)(965,337)
\put(198,337){\makebox(0,0)[r]{32}}
\thicklines \path(220,562)(240,562)
\thicklines \path(985,562)(965,562)
\put(198,562){\makebox(0,0)[r]{36}}
\thicklines \path(220,786)(240,786)
\thicklines \path(985,786)(965,786)
\put(198,786){\makebox(0,0)[r]{40}}
\thicklines \path(220,113)(220,133)
\thicklines \path(220,786)(220,766)
\put(220,68){\makebox(0,0){0}}
\thicklines \path(439,113)(439,133)
\thicklines \path(439,786)(439,766)
\put(439,68){\makebox(0,0){10}}
\thicklines \path(657,113)(657,133)
\thicklines \path(657,786)(657,766)
\put(657,68){\makebox(0,0){20}}
\thicklines \path(876,113)(876,133)
\thicklines \path(876,786)(876,766)
\put(876,68){\makebox(0,0){30}}
\thicklines \path(220,113)(985,113)(985,786)(220,786)(220,113)
\put(45,449){\makebox(0,0)[l]{\shortstack{$\langle\Theta\rangle$}}}
\put(602,0){\makebox(0,0){$\Omega$ [rpm]}}
\put(821,281){\makebox(0,0)[r]{non-MRI}}
\put(865,281){\makebox(0,0){$\star$}}
\put(294,466){\makebox(0,0){$\star$}}
\put(333,506){\makebox(0,0){$\star$}}
\put(376,561){\makebox(0,0){$\star$}}
\put(456,618){\makebox(0,0){$\star$}}
\put(517,673){\makebox(0,0){$\star$}}
\put(584,729){\makebox(0,0){$\star$}}
\put(821,236){\makebox(0,0)[r]{MRI}}
\put(865,236){\circle{18}}
\put(322,225){\circle{18}}
\put(404,281){\circle{18}}
\put(469,336){\circle{18}}
\put(522,393){\circle{18}}
\put(600,450){\circle{18}}
\put(666,506){\circle{18}}
\put(745,561){\circle{18}}
\put(810,616){\circle{18}}
\put(889,673){\circle{18}}
\end{picture}
  \end{center}
  \caption{Comparison of dynamic angle of repose for large mustard seeds 
           taken from MRI ($\circ$) and non-MRI ($\star$) measurements.}
   \label{fig: mri}
\end{figure}
\fi

\section{Simulation Technique}
When measuring the dynamic angle of repose for different materials at the end 
caps, we found that the angle is always lower in the middle of the drum and
the influence of the end caps seems to be rather short range, the angle drops
to the value in the middle of the drum within a few centimeters. We are
particularly interested in the dependence of this length scale on the particle
diameter and density and on the gravitational force. Since we only have a
limited number of particle diameters and density available in the experiment,
we use three-dimensional discrete element methods, also known as {\em granular
dynamics}, to overcome this problem.

Each particle $i$ is approximated by a sphere with radius $r_i$. Only  contact
forces during collisions are considered and the particles are not allowed to
rotate. Since the mustard seeds are slightly aspherical, they rotate much less
than glass beads. This was the motivation for our non-rotating assumption. The
forces acting on particle $i$ during a collision with particle $j$ are 
\begin{equation}
  F_{ij}^n = - \tilde{Y} \ (r_i + r_j - \vec{r}_{ij}\hat{n}) -
               \gamma_n \vec{v}_{ij} \hat{n}
  \label{eq: fn}
\end{equation}
in the normal direction ($\hat{n}$) and
\begin{equation}
  F_{ij}^s = -\min(\gamma_s \vec v_{ij}\cdot\hat{s}(t), \mu|F_{ij}^n|) \ .
  \label{eq: fs_p}
\end{equation}
in the tangential direction ($\hat{s}$) of shearing. In Eqs.~(\ref{eq: fn})
and~(\ref{eq: fs_p}), $\gamma_n$ and $\gamma_s$ represent a dynamic friction
force in the normal and tangential direction, respectively, $\vec{r}_{ij}$
represents the vector  joining both centers of mass, $\vec{v}_{ij}$ represents
the relative motion of  the two particles, and $\tilde{Y}$ is related to the
Young's Modulus of the  investigated material. Dynamic friction in the model is
defined to be proportional to the relative velocity of the particles in the
tangential direction.

During particle--wall contacts, the wall is treated as a particle with infinite
mass and radius. In the normal direction, Eq.~(\ref{eq: fn}) is applied, whereas
in the tangential direction, the static friction force
\begin{equation}
  \tilde{F}_{ij}^s = -\min(k_s \int \vec v_{ij}\cdot\hat{s}(t) dt,
  \mu|F_{ij}^n|)
  \label{eq: fs_w}
\end{equation}
is used. This was motivated by the observation that when particles flow along
the free surface, they dissipate most of their energy in collisions and can
come to rest in voids left by other particles. This is not possible at the flat
drum boundary. In order to avoid additional artificial particles at the walls
which would make the simulations of three-dimensional systems nearly
infeasible, we rather use a static friction law to avoid slipping and allowing
for a static surface angle when the rotation is stopped. Both tangential forces
are limited by the Coulomb criterion which states that the magnitude of the
tangential force cannot exceed the magnitude of the normal force multiplied by
the friction coefficient $\mu$. For particle--particle collisions we use
$\mu=0.2$, and for particle--wall collisions, $\mu_w=0.4$. In order to  save
computer time, we set $\tilde{Y}$ to \mbox{$6\cdot 10^4$ P$\!$a m} which is
about one order of magnitude softer than vulcanite but the maximal overlap of
two particles is at most $0.3\%$ of the sum of their radii, which is still
realistic. This gives a contact time during collisions of $1.1\cdot 10^{-4}$ s.
The coefficient of restitution for wall collisions is set to 0.77 which is
within the error bar of the experimentally measured value of 0.75, see
section~\ref{experiment}. In experiments with spherical liquid-filled
particles, we found only a weak dependence of the restitution coefficient on
particle size and therefore used a normal force law, Eq.~(\ref{eq: fn}), that
would make the restitution coefficient independent of particle size. When the
same type of force law is applied to particle-particle collisions, it gives a
normal restitution coefficient of 0.56. A discussion of the different force
laws is given in ref.~\cite{schaefer96} and a review of different applications
using granular dynamics is given in~\cite{ristow94b}. Even though detailed
experiments for binary collisions of particles were performed, the force
relations before and after a collision depend on the material and the aspherity
of the particles~\cite{foerster94} and since these two quantities were not
available for mustard seeds, we can only take the published values as a
guideline. The numerical parameters were fine adjusted by comparing the
experimentally determined  dynamic angle of repose for 2.5 mm mustard seeds
($\rho = 1.3$ g/cm$^3$) with the simulation results over the $\Omega$--range
from 8 to 35 rpm. The radius of the drum was chosen as $R=3.5$ cm. Both data
sets are shown in Fig.~\ref{fig: fit}. Also shown as a solid line in
Fig.~\ref{fig: fit} is the theoretical result based on a model by Zik et
al.~\cite{zik94}. They started from the equilibrium condition for the surface
flow $j$ in a laminar and thin layer inclined with an angle $\Theta$:
\begin{equation}
  j = \frac{\rho g}{3\eta} h_0^3 \cos\Theta (\tan\Theta - \tan\Theta_0)
  \label{eq: zik1}
\end{equation}
where $\rho$ denotes the particle density, $g$ gravity, $\eta$ the constant
viscosity and $\Theta_0 = \arctan\mu$. The cut-off depth, $h_0$, corresponds to
a constant pressure value of $p_0 = h_0 g \cos\Theta$. A second expression for
the surface flow in a half-filled drum can be obtained by looking at mass
conservation,~\cite{rajchenbach90}:
\begin{equation}
  j = \rho \frac{\Omega}{2} (R^2 - r^2)
  \label{eq: zik2}
\end{equation}
where $r$ measures distance from the drum center along the free surface.
Equating expressions~(\ref{eq: zik1}) and~(\ref{eq: zik2}) and using the 
relation $\tan\Theta=\frac{dy}{dx}=y'$, where $y(x)$ measures the height of the
top surface particle along the surface and $(\cos\Theta)^{-1}=\sqrt{1+(y')^2}$,
we obtain
\begin{equation}
  (y')^3 - (y')^2 \tan\Theta_0 + y' + c \Omega (y^2 + x^2 - R^2) = \tan\Theta_0
  \label{eq: zik3}
\end{equation}
with $c = \frac{3\eta g^2}{2 \rho p_0^3}$. Corrections to this model were
recently proposed by Khakhar et al.~\cite{khakhar97}, but they
lead to the same equations for the dynamic angle of repose as above in the case
of a half-filled drum due to the symmetry of the thickness of the fluidized
layer for shear flow. Solving for $y'$ at the origin (drum center), the only
one of the three roots with no imaginary part reads 
\begin{equation}
  y' = \tilde\mu + (B+\sqrt{D})^{1/3} - \frac{1/3 - \tilde{\mu}}
       {(B+\sqrt{D})^{1/3}}
  \label{eq: angle}
\end{equation}
where $3\tilde\mu = \tan\Theta_0$,
\[
  B = \tilde\mu(1+\tilde\mu^2) + \frac{1}{2}c \Omega R^2 \quad \mbox{and}
\]
\vspace{-5ex}
\[
  D = 3 (\tilde\mu^2 + \frac{1}{9})^2 + \tilde\mu(1+\tilde\mu^2) c \Omega R^2 + 
      (\frac{1}{2}c \Omega R^2)^2 \ .
\] 
We integrated Eq.~(\ref{eq: zik3}) numerically and checked that the
theoretical profile has a similar shape for different rotation speeds as the
numerical data. We adjusted the parameter $c$ using the experimental data
points and the best fit was obtained for $c=0.0111 \pm 0.0001$ s/m$^2$ and
$\mu = 0.51 \Rightarrow \Theta_0 = 27^\circ$. It is remarkable how well the
theoretical curve fits the data points from the experiments and the numerical
simulations, as shown in Fig.~\ref{fig: fit}. The theoretical curve is very
close to the arcus-tangent curve proposed by Hager et al.~\cite{hager97} and
also found in two-dimensional simulations over a wide
$\Omega$-regime~\cite{dury97b}. For low rotation speeds, $\Omega< 8$ rpm, the
experiment is very near the discrete avalanche regime, and therefore the
simulations where we have used only dynamic friction for the particle-particle
interactions and the theory where a steady flow is assumed tend to deviate
slightly from the experiment. Rajchenbach experimentally found the relation
$\Omega \sim |\Theta - \Theta_c|^m$ with $m=0.5$ leading to $\Theta = \Theta_c
+ \alpha \Omega^2$~\cite{rajchenbach90} which gives an increasing slope for
increasing $\Omega$ in the graph, whereas the experimental data points in our
Fig.~\ref{fig: fit} suggest a decreasing slope with increasing $\Omega$. To
illustrate this point further, we replotted all of our available experimental
data points of the large mustard seeds measured at the end caps, taken from
Fig.~\ref{fig: mustard} above, in the same fashion as Rajchenbach and obtained
a scaling exponent of m=0.87 using $\Theta_c=34.1^\circ$. This has to be
compared with m=0.5 found by Rajchenbach and m=0.7 given by the numerical
prediction by Tang and Bak~\cite{tang88}. The $\Omega$ range of Rajchenbach is
smaller than the one investigated by us and we speculate that his finding is
valid close to the transition point to the continuous flow regime where the
quadratic fit works rather well. We included this in Fig.~\ref{fig: fit} as
dotted line using a best quadratic fit for the value of $\Omega<20$ rpm. 
\ifx\grdraft\undefined
\else
\vspace{-4ex}
\begin{figure}[htb]
  \begin{center}
\setlength{\unitlength}{0.240900pt}
\begin{picture}(1049,809)(0,0)
\thicklines \path(220,113)(240,113)
\thicklines \path(985,113)(965,113)
\put(198,113){\makebox(0,0)[r]{24}}
\thicklines \path(220,281)(240,281)
\thicklines \path(985,281)(965,281)
\put(198,281){\makebox(0,0)[r]{28}}
\thicklines \path(220,450)(240,450)
\thicklines \path(985,450)(965,450)
\put(198,450){\makebox(0,0)[r]{32}}
\thicklines \path(220,618)(240,618)
\thicklines \path(985,618)(965,618)
\put(198,618){\makebox(0,0)[r]{36}}
\thicklines \path(220,786)(240,786)
\thicklines \path(985,786)(965,786)
\put(198,786){\makebox(0,0)[r]{40}}
\thicklines \path(220,113)(220,133)
\thicklines \path(220,786)(220,766)
\put(220,68){\makebox(0,0){0}}
\thicklines \path(439,113)(439,133)
\thicklines \path(439,786)(439,766)
\put(439,68){\makebox(0,0){10}}
\thicklines \path(657,113)(657,133)
\thicklines \path(657,786)(657,766)
\put(657,68){\makebox(0,0){20}}
\thicklines \path(876,113)(876,133)
\thicklines \path(876,786)(876,766)
\put(876,68){\makebox(0,0){30}}
\thicklines \path(220,113)(985,113)(985,786)(220,786)(220,113)
\put(45,449){\makebox(0,0)[l]{\shortstack{$\langle\Theta\rangle$}}}
\put(602,0){\makebox(0,0){$\Omega$ [rpm]}}
\put(854,281){\makebox(0,0)[r]{simulation}}
\put(898,281){\circle*{12}}
\put(293,271){( \circle*{12} )}
\put(377,357){\circle*{12}}
\put(429,424){\circle*{12}}
\put(533,511){\circle*{12}}
\put(637,576){\circle*{12}}
\put(742,630){\circle*{12}}
\put(846,676){\circle*{12}}
\put(898,696){\circle*{12}}
\thinlines \path(876,281)(942,281)
\thinlines \path(876,291)(876,271)
\thinlines \path(942,291)(942,271)
\thinlines \path(324,266)(324,276)
\thinlines \path(314,266)(334,266)
\thinlines \path(314,276)(334,276)
\thinlines \path(377,354)(377,360)
\thinlines \path(367,354)(387,354)
\thinlines \path(367,360)(387,360)
\thinlines \path(429,420)(429,428)
\thinlines \path(419,420)(439,420)
\thinlines \path(419,428)(439,428)
\thinlines \path(533,508)(533,514)
\thinlines \path(523,508)(543,508)
\thinlines \path(523,514)(543,514)
\thinlines \path(637,573)(637,580)
\thinlines \path(627,573)(647,573)
\thinlines \path(627,580)(647,580)
\thinlines \path(742,627)(742,634)
\thinlines \path(732,627)(752,627)
\thinlines \path(732,634)(752,634)
\thinlines \path(846,672)(846,680)
\thinlines \path(836,672)(856,672)
\thinlines \path(836,680)(856,680)
\thinlines \path(898,692)(898,699)
\thinlines \path(888,692)(908,692)
\thinlines \path(888,699)(908,699)
\put(854,236){\makebox(0,0)[r]{experiment}}
\put(898,236){\circle{18}}
\put(322,365){\circle{18}}
\put(404,407){\circle{18}}
\put(469,449){\circle{18}}
\put(522,491){\circle{18}}
\put(600,534){\circle{18}}
\put(666,576){\circle{18}}
\put(745,617){\circle{18}}
\put(810,658){\circle{18}}
\put(889,701){\circle{18}}
\put(854,191){\makebox(0,0)[r]{theory}}
\thinlines \path(876,191)(942,191)
\thinlines \path(325,344)(325,344)(431,427)(536,499)(642,565)(747,625)(852,680)(958,732)
\put(854,146){\makebox(0,0)[r]{quadratic fit}}
\dottedline{12}(876,146)(942,146)
\dottedline{12}(313,371)(320,373)(328,375)(336,378)(344,380)(351,383)(359,386)(367,390)(375,393)(382,397)(390,400)(398,404)(405,408)(413,413)(421,417)(429,422)(436,426)(444,431)(452,437)(460,442)(467,447)(475,453)(483,459)(490,465)(498,471)(506,478)(514,484)(521,491)(529,498)(537,505)(545,512)(552,519)(560,527)(568,535)(575,543)(583,551)(591,559)(599,568)
\dottedline{12}(599,568)(606,576)(614,585)(622,594)(630,603)(637,612)(645,622)(653,632)(660,641)(668,652)(676,662)(684,672)(691,683)(699,693)(707,704)(715,715)(722,727)(730,738)(738,750)(745,761)(753,773)(761,786)(761,786)
\end{picture}
  \end{center}
  \caption[]{Comparison of dynamic angle of repose for large mustard seeds 
	   taken from MRI ($\circ$), numerical simulation ($\bullet$) and the
	   theory of Zik et al.~\cite{zik94} (---).}
   \label{fig: fit}
\end{figure}
\fi

\section{Boundary Effect on Surface Angle}
As shown in Fig.~\ref{fig: mri}, the dynamic angle of repose of granular
material in a rotating drum is significantly higher at the end caps than in the
middle. For the experimentally investigated particle sizes of the order of 
millimeter, the effect was visible up to a few centimeters. But this length
scale might depend on the particle diameter and density and also on external
parameters like gravity.

Using the above described technique, we simulated extended three-dimensional,
half-filled drums. For a particle size of 2.5 mm and a rotation speed of 20
rpm, the time-averaged angle, denoted by $\langle\Theta(z)\rangle$, as function
of the position along the rotation axis is shown in Fig.~\ref{fig: profile}.
Each point with corresponding error bar stands for a weighted average over the
nearest neighbours. In order to study the characteristic length, $\zeta$, of
the boundary effect, we fit all data points by the relation
\begin{equation}
  \langle\Theta(z)\rangle = \Theta_\infty + \Delta\Theta\left( 
                            e^{-z/\zeta} + e^{-(L-z)/\zeta} \right)
  \label{eq: profile}
\end{equation}
where $L$ stands for the length of the drum, $\Theta_\infty$ for the dynamic
angle of repose far away from the boundaries and $\Delta\Theta$ for the angle
difference between the value at the boundary and $\Theta_\infty$. For the curve
shown in Fig.~\ref{fig: profile}, the corresponding values are
$\Delta\Theta\approx 4^\circ$, which is the same value given in Fig.~\ref{fig:
mri} for the MRI experiment, $\zeta = 3.19 \pm 0.25$, $L = 20.6$ cm and
$\Theta_\infty = 35^\circ \pm 0.2^\circ$. We tested Eq.~(\ref{eq: profile})
against the simulation results for different drum lengths of $L/2$, $L/4$ and 
$L/8$ and found a remarkably good  agreement. In the last two cases, the value
for $\Theta_\infty$ is never reached in the middle of the drum  due to the
boundary effects.
\ifx\grdraft\undefined
\else
\vspace{-4ex}
\begin{figure}[htb]
  \begin{center}
    \input bound_6
  \end{center}
  \caption{Profile of the dynamic angle of repose along the rotation axis 
           for 2.5 mm spheres: ($\bullet$) simulation, (---) fit.}
   \label{fig: profile}
\end{figure}
\fi

Using drums that are at least 64 particle diameters long, we studied the
dependence of $\Theta_\infty$ (middle) and the angle at the end caps on the
particle diameter $d$ at the same rotation speed of 20 rpm. The simulation
values in the tangential direction were chosen in such a way that the
normalized tangential velocities before and after impact were {\em independent}
of the particle diameter $d$. The ration $\tilde{Y}/k_s$ was set to 3.1, a
value which gives for acetate spheres a good agreement of
simulations~\cite{schaefer96} and experiments~\cite{foerster94}. The value of
$\gamma_s$ was chosen sufficiently high to give a similar behavior in
particle-particle- and particle-wall-collisions in the sliding regime. The
results for the surface angle along the rotation axis are shown in
Fig.~\ref{fig: angle}, which illustrates that the angle increases with
increasing particle size in agreement with the mustard seed experimental
results given in Fig.~\ref{fig: mustard}. The angle difference of around 4
degrees, which seems to be independent of the particle size, also agrees with
the experimental findings. In other experiments, different dependencies were
observed: das Gupta et al.~\cite{dasgupta91} mostly found a higher angle for
smaller particles using sand grains and Hill and Kakalios~\cite{hill94}
measured higher angles for smaller particles when using sand and glass
particles although it was also possible to get no angle difference for certain
size ratios of glass spheres. The latter was also found by Zik et
al.~\cite{zik94}, whereas Cantelaube~\cite{cantelaube95b} did not find a clear
trend when using discs in a quasi two-dimensional drum. What causes the
different behaviour is not clear at the moment and a more detailed analysis
would be desirable but is beyond the scope of this article. It is necessary to
use appropriate values for the  simulation parameters to quantitatively model a
desired system, which is why we gathered as much information for the mustard
seeds as possible. A arcus-tangent fit which gave a smaller mean deviation than
a parabolic fit was added to Fig.~\ref{fig: angle} to guide the eye. Changing
the density, $\rho$, of the particles or the gravitational constant, $g$, has a
dramatic effect on the angle of repose: for the latter quantity this is shown
in table~\ref{tab: gravity} and a similar behaviour was seen in recent
experiments~\cite{fabi97b}. For both $\rho$ and $g$, an increase in value
corresponds to an angle decrease. When a hydrostatic pressure, $p_0 \sim g$, is
assumed, the data for $g<30$ m/s$^2$ can be well described by Eq.~(\ref{eq:
angle}). The lower $g$ becomes the more pronounced the S-shaped surface
becomes, and in the limit $g \rightarrow 0$, the transition to the centrifugal
regime takes place at
\[
  \Omega_c \approx \sqrt{\frac{\sqrt{2}\,g}{R\,\sin\Theta_0}}
\]
where $\Theta_0$ denotes the average angle in the limit $\Omega \rightarrow
0$~\cite{walton93,dury97b}. In our case, for $\Omega = 20$ rpm the transition
to the centrifugal regime occurs at $g \approx 0.45$ m/s$^2$, which is in
perfect agreement with the numerical findings. Even though we studied more than
one (two) orders of magnitude in $\rho$ ($g$), we could not obtain an accurate
infinite value limit.
\ifx\grdraft\undefined
\else
\vspace{-4ex}
\begin{figure}[htb]
  \begin{center}
\setlength{\unitlength}{0.240900pt}
\begin{picture}(1049,809)(0,0)
\thicklines \path(220,113)(240,113)
\thicklines \path(985,113)(965,113)
\put(198,113){\makebox(0,0)[r]{26}}
\thicklines \path(220,248)(240,248)
\thicklines \path(985,248)(965,248)
\put(198,248){\makebox(0,0)[r]{30}}
\thicklines \path(220,382)(240,382)
\thicklines \path(985,382)(965,382)
\put(198,382){\makebox(0,0)[r]{34}}
\thicklines \path(220,517)(240,517)
\thicklines \path(985,517)(965,517)
\put(198,517){\makebox(0,0)[r]{38}}
\thicklines \path(220,651)(240,651)
\thicklines \path(985,651)(965,651)
\put(198,651){\makebox(0,0)[r]{42}}
\thicklines \path(220,786)(240,786)
\thicklines \path(985,786)(965,786)
\put(198,786){\makebox(0,0)[r]{46}}
\thicklines \path(220,113)(220,133)
\thicklines \path(220,786)(220,766)
\put(220,68){\makebox(0,0){0}}
\thicklines \path(411,113)(411,133)
\thicklines \path(411,786)(411,766)
\put(411,68){\makebox(0,0){2}}
\thicklines \path(603,113)(603,133)
\thicklines \path(603,786)(603,766)
\put(603,68){\makebox(0,0){4}}
\thicklines \path(794,113)(794,133)
\thicklines \path(794,786)(794,766)
\put(794,68){\makebox(0,0){6}}
\thicklines \path(985,113)(985,133)
\thicklines \path(985,786)(985,766)
\put(985,68){\makebox(0,0){8}}
\thicklines \path(220,113)(985,113)(985,786)(220,786)(220,113)
\put(45,449){\makebox(0,0)[l]{\shortstack{$\langle\Theta\rangle$}}}
\put(602,23){\makebox(0,0){$d$ [mm]}}
\put(842,248){\makebox(0,0)[r]{end cap}}
\put(886,248){\makebox(0,0){$\star$}}
\put(889,749){\makebox(0,0){$\star$}}
\put(794,715){\makebox(0,0){$\star$}}
\put(698,688){\makebox(0,0){$\star$}}
\put(603,655){\makebox(0,0){$\star$}}
\put(555,624){\makebox(0,0){$\star$}}
\put(507,591){\makebox(0,0){$\star$}}
\put(459,571){\makebox(0,0){$\star$}}
\put(411,537){\makebox(0,0){$\star$}}
\put(363,490){\makebox(0,0){$\star$}}
\put(316,416){\makebox(0,0){$\star$}}
\thinlines \path(864,248)(930,248)
\thinlines \path(864,258)(864,238)
\thinlines \path(930,258)(930,238)
\thinlines \path(889,730)(889,767)
\thinlines \path(879,730)(899,730)
\thinlines \path(879,767)(899,767)
\thinlines \path(794,697)(794,734)
\thinlines \path(784,697)(804,697)
\thinlines \path(784,734)(804,734)
\thinlines \path(698,673)(698,704)
\thinlines \path(688,673)(708,673)
\thinlines \path(688,704)(708,704)
\thinlines \path(603,640)(603,670)
\thinlines \path(593,640)(613,640)
\thinlines \path(593,670)(613,670)
\thinlines \path(555,614)(555,635)
\thinlines \path(545,614)(565,614)
\thinlines \path(545,635)(565,635)
\thinlines \path(507,576)(507,606)
\thinlines \path(497,576)(517,576)
\thinlines \path(497,606)(517,606)
\thinlines \path(459,564)(459,577)
\thinlines \path(449,564)(469,564)
\thinlines \path(449,577)(469,577)
\thinlines \path(411,529)(411,545)
\thinlines \path(401,529)(421,529)
\thinlines \path(401,545)(421,545)
\thinlines \path(363,483)(363,497)
\thinlines \path(353,483)(373,483)
\thinlines \path(353,497)(373,497)
\thinlines \path(316,409)(316,423)
\thinlines \path(306,409)(326,409)
\thinlines \path(306,423)(326,423)
\put(842,203){\makebox(0,0)[r]{middle}}
\put(886,203){\circle*{12}}
\put(889,645){\circle*{12}}
\put(794,608){\circle*{12}}
\put(698,567){\circle*{12}}
\put(603,517){\circle*{12}}
\put(555,480){\circle*{12}}
\put(507,453){\circle*{12}}
\put(459,416){\circle*{12}}
\put(411,369){\circle*{12}}
\put(363,315){\circle*{12}}
\put(316,238){\circle*{12}}
\thinlines \path(864,203)(930,203)
\thinlines \path(864,213)(864,193)
\thinlines \path(930,213)(930,193)
\thinlines \path(889,635)(889,655)
\thinlines \path(879,635)(899,635)
\thinlines \path(879,655)(899,655)
\thinlines \path(794,594)(794,621)
\thinlines \path(784,594)(804,594)
\thinlines \path(784,621)(804,621)
\thinlines \path(698,557)(698,577)
\thinlines \path(688,557)(708,557)
\thinlines \path(688,577)(708,577)
\thinlines \path(603,507)(603,527)
\thinlines \path(593,507)(613,507)
\thinlines \path(593,527)(613,527)
\thinlines \path(555,473)(555,487)
\thinlines \path(545,473)(565,473)
\thinlines \path(545,487)(565,487)
\thinlines \path(507,439)(507,466)
\thinlines \path(497,439)(517,439)
\thinlines \path(497,466)(517,466)
\thinlines \path(459,412)(459,419)
\thinlines \path(449,412)(469,412)
\thinlines \path(449,419)(469,419)
\thinlines \path(411,362)(411,375)
\thinlines \path(401,362)(421,362)
\thinlines \path(401,375)(421,375)
\thinlines \path(363,308)(363,322)
\thinlines \path(353,308)(373,308)
\thinlines \path(353,322)(373,322)
\thinlines \path(316,231)(316,244)
\thinlines \path(306,231)(326,231)
\thinlines \path(306,244)(326,244)
\thinlines \path(228,113)(235,124)(243,136)(251,148)(259,160)(266,172)(274,184)(282,195)(290,207)(297,218)(305,229)(313,240)(320,251)(328,262)(336,272)(344,283)(351,293)(359,303)(367,313)(375,322)(382,332)(390,341)(398,350)(405,359)(413,367)(421,375)(429,384)(436,392)(444,399)(452,407)(460,414)(467,422)(475,429)(483,435)(490,442)(498,449)(506,455)(514,461)(521,467)(529,473)(537,479)(545,484)(552,490)(560,495)(568,500)(575,505)(583,510)(591,515)(599,520)(606,524)(614,528)
\thinlines \path(614,528)(622,533)(630,537)(637,541)(645,545)(653,549)(660,553)(668,557)(676,560)(684,564)(691,567)(699,571)(707,574)(715,577)(722,580)(730,583)(738,586)(745,589)(753,592)(761,595)(769,598)(776,601)(784,603)(792,606)(800,608)(807,611)(815,613)(823,616)(830,618)(838,620)(846,623)(854,625)(861,627)(869,629)(877,631)(885,633)(892,635)(900,637)(908,639)(915,641)(923,643)(931,644)(939,646)(946,648)(954,650)(962,651)(970,653)(977,655)(985,656)
\thinlines \path(220,274)(220,274)(228,287)(235,299)(243,312)(251,324)(259,336)(266,348)(274,360)(282,372)(290,384)(297,395)(305,406)(313,417)(320,428)(328,438)(336,448)(344,458)(351,467)(359,477)(367,486)(375,494)(382,503)(390,511)(398,519)(405,526)(413,534)(421,541)(429,548)(436,554)(444,561)(452,567)(460,573)(467,579)(475,585)(483,590)(490,595)(498,600)(506,605)(514,610)(521,615)(529,619)(537,623)(545,627)(552,632)(560,635)(568,639)(575,643)(583,646)(591,650)(599,653)
\thinlines \path(599,653)(606,656)(614,660)(622,663)(630,666)(637,669)(645,671)(653,674)(660,677)(668,679)(676,682)(684,684)(691,687)(699,689)(707,691)(715,693)(722,696)(730,698)(738,700)(745,702)(753,704)(761,706)(769,707)(776,709)(784,711)(792,713)(800,714)(807,716)(815,718)(823,719)(830,721)(838,722)(846,724)(854,725)(861,726)(869,728)(877,729)(885,730)(892,732)(900,733)(908,734)(915,735)(923,737)(931,738)(939,739)(946,740)(954,741)(962,742)(970,743)(977,744)(985,745)
\end{picture}
  \end{center}
  \caption{Dynamic angle of repose as function of sphere diameter for 
           $\Omega=20$ rpm (simulation); ($\star$) end cap, ($\bullet$) drum 
           middle, (---) arcus-tangent fit.}
   \label{fig: angle}
\end{figure}
\begin{table}[htb]
  \begin{center}
    \begin{tabular}{l||c|c|c|c|c|c}
    $g$ [m/s$^2$]       & 1.62 & 3.73 & 9.81 & 13.6 & 25.1 &  274 \\ \hline
    $\Theta$ [$^\circ$] & 48.2 & 41.4 & 35.0 & 33.3 & 30.2 & 18.4 \\ \hline
    $\zeta/R$           &0.291 &0.274 &0.277 & 0.269& 0.283& 0.260
    \end{tabular}
  \end{center}
  \caption{Angle of repose, $\Theta$, in the drum middle and dimensionless 
           characteristic length, $\zeta/R$, as function of gravity, $g$ 
           (simulation).}
  \label{tab: gravity}
\end{table}
\fi

\section{Range of Boundary Effect}
In order to study the range of the boundary effect, we extracted figures
similar to Fig.~\ref{fig: profile} from our simulations and varied the drum
length, $L$, and radius, $R$, the particle diameter, $d$, and density, $\rho$,
the gravitational constant, $g$, and the rotation speed, $\Omega$. The data
points for the dynamic angle of repose as a function of position along the
rotation axis were fitted by Eq.~(\ref{eq: profile}) giving the characteristic
length, $\zeta$, of this run. As expected, $\zeta$ did not vary when the length
of the drum or the rotation speed was changed, but surprisingly, the
characteristic length, $\zeta$, did not change when the density of the
particles or the gravitational constant was changed by more than one order of
magnitude, even though the dynamic angle of repose strongly depends on both as
shown in table~\ref{tab: gravity} for the latter quantity.
\ifx\grdraft\undefined
\else
\vspace{-4ex}
\begin{figure}[htb]
  \begin{center}
\setlength{\unitlength}{0.240900pt}
\begin{picture}(1500,900)(0,0)
\thicklines \path(199,179)(219,179)
\thicklines \path(952,179)(932,179)
\put(177,179){\makebox(0,0)[r]{0}}
\thicklines \path(199,292)(219,292)
\thicklines \path(952,292)(932,292)
\put(177,292){\makebox(0,0)[r]{}}
\thicklines \path(199,404)(219,404)
\thicklines \path(952,404)(932,404)
\put(177,404){\makebox(0,0)[r]{0.2}}
\thicklines \path(199,517)(219,517)
\thicklines \path(952,517)(932,517)
\put(177,517){\makebox(0,0)[r]{}}
\thicklines \path(199,630)(219,630)
\thicklines \path(952,630)(932,630)
\put(177,630){\makebox(0,0)[r]{0.4}}
\thicklines \path(199,742)(219,742)
\thicklines \path(952,742)(932,742)
\put(177,742){\makebox(0,0)[r]{}}
\thicklines \path(199,855)(219,855)
\thicklines \path(952,855)(932,855)
\put(177,855){\makebox(0,0)[r]{0.6}}
\thicklines \path(199,179)(199,199)
\thicklines \path(199,855)(199,835)
\put(199,134){\makebox(0,0){0}}
\thicklines \path(378,179)(378,199)
\thicklines \path(378,855)(378,835)
\put(378,134){\makebox(0,0){0.05}}
\thicklines \path(558,179)(558,199)
\thicklines \path(558,855)(558,835)
\put(558,134){\makebox(0,0){0.1}}
\thicklines \path(737,179)(737,199)
\thicklines \path(737,855)(737,835)
\put(737,134){\makebox(0,0){0.15}}
\thicklines \path(916,179)(916,199)
\thicklines \path(916,855)(916,835)
\put(916,134){\makebox(0,0){0.2}}
\thicklines \path(199,179)(952,179)(952,855)(199,855)(199,179)
\put(45,517){\makebox(0,0)[l]{\shortstack{$\zeta/R$}}}
\put(575,67){\makebox(0,0){$d/R$}}
\put(805,348){\makebox(0,0)[r]{$R=3.5$ cm}}
\thinlines \path(827,348)(935,348)
\thinlines \path(827,358)(827,338)
\thinlines \path(935,358)(935,338)
\thinlines \path(916,610)(916,791)
\thinlines \path(906,610)(926,610)
\thinlines \path(906,791)(926,791)
\thinlines \path(814,549)(814,659)
\thinlines \path(804,549)(824,549)
\thinlines \path(804,659)(824,659)
\thinlines \path(711,453)(711,549)
\thinlines \path(701,453)(721,453)
\thinlines \path(701,549)(721,549)
\thinlines \path(660,443)(660,507)
\thinlines \path(650,443)(670,443)
\thinlines \path(650,507)(670,507)
\thinlines \path(609,449)(609,527)
\thinlines \path(599,449)(619,449)
\thinlines \path(599,527)(619,527)
\thinlines \path(558,465)(558,517)
\thinlines \path(548,465)(568,465)
\thinlines \path(548,517)(568,517)
\thinlines \path(506,443)(506,520)
\thinlines \path(496,443)(516,443)
\thinlines \path(496,520)(516,520)
\thinlines \path(455,472)(455,511)
\thinlines \path(445,472)(465,472)
\thinlines \path(445,511)(465,511)
\thinlines \path(404,475)(404,514)
\thinlines \path(394,475)(414,475)
\thinlines \path(394,514)(414,514)
\thinlines \path(353,485)(353,517)
\thinlines \path(343,485)(363,485)
\thinlines \path(343,517)(363,517)
\thinlines \path(301,533)(301,559)
\thinlines \path(291,533)(311,533)
\thinlines \path(291,559)(311,559)
\put(916,700){\makebox(0,0){$\star$}}
\put(814,604){\makebox(0,0){$\star$}}
\put(711,501){\makebox(0,0){$\star$}}
\put(660,475){\makebox(0,0){$\star$}}
\put(609,488){\makebox(0,0){$\star$}}
\put(558,491){\makebox(0,0){$\star$}}
\put(506,482){\makebox(0,0){$\star$}}
\put(455,491){\makebox(0,0){$\star$}}
\put(404,494){\makebox(0,0){$\star$}}
\put(353,501){\makebox(0,0){$\star$}}
\put(301,546){\makebox(0,0){$\star$}}
\put(881,348){\makebox(0,0){$\star$}}
\put(805,303){\makebox(0,0)[r]{$7$ cm}}
\put(353,504){\circle{18}}
\put(404,512){\circle{18}}
\put(506,503){\circle{18}}
\put(609,519){\circle{18}}
\put(711,533){\circle{18}}
\put(814,628){\circle{18}}
\put(916,662){\circle{18}}
\put(881,303){\circle{18}}
\put(805,258){\makebox(0,0)[r]{$10.5$ cm}}
\put(404,503){\circle*{12}}
\put(506,506){\circle*{12}}
\put(609,555){\circle*{12}}
\put(881,258){\circle*{12}}
\thinlines \drawline[-50](199,494)(199,494)(207,494)(214,494)(222,494)(229,494)(237,494)(245,494)(252,494)(260,494)(267,494)(275,494)(283,494)(290,494)(298,494)(305,494)(313,494)(321,494)(328,494)(336,494)(344,494)(351,494)(359,494)(366,494)(374,494)(382,494)(389,494)(397,494)(404,494)(412,494)(420,494)(427,494)(435,494)(442,494)(450,494)(458,494)(465,494)(473,494)(480,494)(488,494)(496,494)(503,494)(511,494)(518,494)(526,494)(534,494)(541,494)(549,494)(556,494)(564,494)(572,494)
\thinlines \drawline[-50](572,494)(579,494)(587,494)(595,494)(602,494)(610,494)(617,494)(625,494)(633,494)(640,494)(648,494)(655,494)(663,494)(671,494)(678,494)(686,494)(693,494)(701,494)(709,494)(716,494)(724,494)(731,494)(739,494)(747,494)(754,494)(762,494)(769,494)(777,494)(785,494)(792,494)(800,494)(807,494)(815,494)(823,494)(830,494)(838,494)(846,494)(853,494)(861,494)(868,494)(876,494)(884,494)(891,494)(899,494)(906,494)(914,494)(922,494)(929,494)(937,494)(944,494)(952,494)
\end{picture}
  \end{center}
  \caption{Dimensionless range of boundary effect for spheres with different 
           diameter; ($\star$) $R=3.5$ cm, ($\circ$) $R=7$ cm, ($\bullet$)
           $R=10.5$ cm, dotted line shows value $\zeta=0.28 \, R$ for 
           radius-independent regime (simulation).}
   \label{fig: range}
\end{figure}
\fi

Based on the definition of $\zeta$ in Eq.~(\ref{eq: profile}), one might
speculate that $\zeta \sim R$ since the gradient of the slope along the
rotational axis of the surface should be a material property, i.e.\ it should
not depend on the geometry. The angle of repose is independent of the drum
radius, $R$, and therefore the height difference between the surface at the end
cap and the surface in the middle of the drum must be proportional to $R$. This
leads to $\zeta \sim R$ which is indeed the case, and we show in Fig.~\ref{fig:
range} the dimensionless characteristic length, $\zeta/R$, as function of
dimensionless particle diameter, $d/R$, for three different drum radii. Below a
critical diameter, $d_c$, $\zeta$ seems to be independent of the particle size,
and we propose the following relation
\begin{equation}
  \zeta = \left\{ \begin{array}{ll}
                      \alpha\, R & \mbox{, if $d \le d_c$} \\
                      \alpha\, R + \beta (d - d_c) & \mbox{, if $d>d_c$}
                  \end{array}
          \right.
\end{equation}
where, in our case, $\alpha=0.28$ and $\beta= 3.13$. The critical particle
diameter $d_c \approx 0.14\, R$ and it seems to decrease slightly with
increasing drum radius. Therefore, particles in the fluidized zone with $d<d_c$
might be describable by a continuum model. For particles with $d>d_c$, we have 
to take finite size effects into account.

For comparison, we replot in Fig.~\ref{fig: range3} the data from
Fig.~\ref{fig: range} by showing the characteristic length, $\zeta$, made
dimensionless by the average particle diameter, $d$. A strong decrease in
$\zeta/d$ and a later saturation is clearly visible for increasing particle
diameter. The solid line is an exponential fit which matches all data points
very well but it can  only serve as a guide to the eye since it does not
reproduce the right value in the limit $d \rightarrow 0$.
\ifx\grdraft\undefined
\else
\vspace{-4ex}
\begin{figure}[htb]
  \begin{center}
\setlength{\unitlength}{0.240900pt}
\begin{picture}(1500,900)(0,0)
\thicklines \path(199,179)(219,179)
\thicklines \path(952,179)(932,179)
\put(177,179){\makebox(0,0)[r]{0}}
\thicklines \path(199,292)(219,292)
\thicklines \path(952,292)(932,292)
\put(177,292){\makebox(0,0)[r]{}}
\thicklines \path(199,404)(219,404)
\thicklines \path(952,404)(932,404)
\put(177,404){\makebox(0,0)[r]{0.4}}
\thicklines \path(199,517)(219,517)
\thicklines \path(952,517)(932,517)
\put(177,517){\makebox(0,0)[r]{}}
\thicklines \path(199,630)(219,630)
\thicklines \path(952,630)(932,630)
\put(177,630){\makebox(0,0)[r]{0.8}}
\thicklines \path(199,742)(219,742)
\thicklines \path(952,742)(932,742)
\put(177,742){\makebox(0,0)[r]{}}
\thicklines \path(199,855)(219,855)
\thicklines \path(952,855)(932,855)
\put(177,855){\makebox(0,0)[r]{1.2}}
\thicklines \path(199,179)(199,199)
\thicklines \path(199,855)(199,835)
\put(199,134){\makebox(0,0){0}}
\thicklines \path(378,179)(378,199)
\thicklines \path(378,855)(378,835)
\put(378,134){\makebox(0,0){0.05}}
\thicklines \path(558,179)(558,199)
\thicklines \path(558,855)(558,835)
\put(558,134){\makebox(0,0){0.1}}
\thicklines \path(737,179)(737,199)
\thicklines \path(737,855)(737,835)
\put(737,134){\makebox(0,0){0.15}}
\thicklines \path(916,179)(916,199)
\thicklines \path(916,855)(916,835)
\put(916,134){\makebox(0,0){0.2}}
\thicklines \path(199,179)(952,179)(952,855)(199,855)(199,179)
\put(45,517){\makebox(0,0)[l]{\shortstack{$\zeta/d$}}}
\put(575,67){\makebox(0,0){$d/R$}}
\put(805,742){\makebox(0,0)[r]{$R=3.5$ cm}}
\thinlines \path(827,742)(935,742)
\thinlines \path(827,752)(827,732)
\thinlines \path(935,752)(935,732)
\thinlines \path(916,287)(916,332)
\thinlines \path(906,287)(926,287)
\thinlines \path(906,332)(926,332)
\thinlines \path(814,287)(814,319)
\thinlines \path(804,287)(824,287)
\thinlines \path(804,319)(824,319)
\thinlines \path(711,275)(711,309)
\thinlines \path(701,275)(721,275)
\thinlines \path(701,309)(721,309)
\thinlines \path(660,282)(660,307)
\thinlines \path(650,282)(670,282)
\thinlines \path(650,307)(670,307)
\thinlines \path(609,297)(609,331)
\thinlines \path(599,297)(619,297)
\thinlines \path(599,331)(619,331)
\thinlines \path(558,322)(558,348)
\thinlines \path(548,322)(568,322)
\thinlines \path(548,348)(568,348)
\thinlines \path(506,333)(506,378)
\thinlines \path(496,333)(516,333)
\thinlines \path(496,378)(516,378)
\thinlines \path(455,384)(455,411)
\thinlines \path(445,384)(465,384)
\thinlines \path(445,411)(465,411)
\thinlines \path(404,438)(404,472)
\thinlines \path(394,438)(414,438)
\thinlines \path(394,472)(414,472)
\thinlines \path(353,536)(353,573)
\thinlines \path(343,536)(363,536)
\thinlines \path(343,573)(363,573)
\thinlines \path(301,799)(301,844)
\thinlines \path(291,799)(311,799)
\thinlines \path(291,844)(311,844)
\put(916,309){\makebox(0,0){$\star$}}
\put(814,303){\makebox(0,0){$\star$}}
\put(711,292){\makebox(0,0){$\star$}}
\put(660,294){\makebox(0,0){$\star$}}
\put(609,314){\makebox(0,0){$\star$}}
\put(558,335){\makebox(0,0){$\star$}}
\put(506,356){\makebox(0,0){$\star$}}
\put(455,398){\makebox(0,0){$\star$}}
\put(404,455){\makebox(0,0){$\star$}}
\put(353,555){\makebox(0,0){$\star$}}
\put(301,821){\makebox(0,0){$\star$}}
\put(881,742){\makebox(0,0){$\star$}}
\put(805,697){\makebox(0,0)[r]{$7$ cm}}
\put(353,558){\circle{18}}
\put(404,471){\circle{18}}
\put(506,368){\circle{18}}
\put(609,328){\circle{18}}
\put(711,303){\circle{18}}
\put(814,310){\circle{18}}
\put(916,300){\circle{18}}
\put(881,697){\circle{18}}
\put(805,652){\makebox(0,0)[r]{$10.5$ cm}}
\put(404,463){\circle*{12}}
\put(506,370){\circle*{12}}
\put(609,343){\circle*{12}}
\put(881,652){\circle*{12}}
\put(805,607){\makebox(0,0)[r]{exp-fit}}
\thinlines \path(827,607)(935,607)
\thinlines \path(294,855)(298,828)(305,784)(313,743)(321,706)(328,672)(336,641)(344,612)(351,586)(359,562)(366,540)(374,520)(382,501)(389,485)(397,469)(404,455)(412,442)(420,430)(427,419)(435,409)(442,400)(450,391)(458,383)(465,376)(473,370)(480,364)(488,358)(496,353)(503,349)(511,345)(518,341)(526,337)(534,334)(541,331)(549,328)(556,326)(564,324)(572,322)(579,320)(587,318)(595,316)(602,315)(610,313)(617,312)(625,311)(633,310)(640,309)(648,308)(655,307)(663,307)(671,306)
\thinlines \path(671,306)(678,305)(686,305)(693,304)(701,304)(709,303)(716,303)(724,303)(731,302)(739,302)(747,302)(754,301)(762,301)(769,301)(777,301)(785,301)(792,300)(800,300)(807,300)(815,300)(823,300)(830,300)(838,300)(846,300)(853,300)(861,300)(868,299)(876,299)(884,299)(891,299)(899,299)(906,299)(914,299)(922,299)(929,299)(937,299)(944,299)(952,299)
\end{picture}
  \end{center}
  \caption{Range of boundary effect measured in sphere diameters for spheres 
           with different diameter; ($\star$) $R=3.5$ cm, ($\circ$) $R=7$ cm,
           ($\bullet$) $R=10.5$ cm, (---) exponential fit (simulation).}
   \label{fig: range3}
\end{figure}
\fi

\section{Conclusions}
We have investigated the dynamic angle of repose, $\Theta$, in a
three-dimensional rotating drum in the continuous flow regime. By choosing
different materials and particle diameters, we discussed the
$\Omega$-dependence of $\Theta$ for glass  beads and mustard seeds of two
different sizes. In the low rotation speed regime,  both types of glass beads
showed the same angle of repose, whereas the angle was higher for the {\em
larger} mustard seeds. Using MRI techniques, we could quantify, for the large
mustard seeds, the angle difference between the middle and the end of the drum
and its $\Omega$-dependence. In all cases, either a linear or an arcus-tangent
dependence of $\Theta$ on $\Omega$ was found. In order to investigate the
range, $\zeta$, of the boundary effects, we used a three-dimensional {\em
discrete element} code and fitted the averaged angle along the rotation axis to
two exponentially decaying functions. We found that $\zeta$ scales linearly
with the drum radius. On the other hand, it does not depend either on the
particle  density or the gravitational constant, even though the surface angle
changes drastically with these quantities, or on the rotation speed of the
drum. A detailed analysis of the dependence of the characteristic length,
$\zeta$, on the particle diameter, $d$, revealed  that $\zeta$ is independent
of $d$ for small particle diameters but shows finite size effects for larger
$d$.

\acknowledgments
CMD and GHR gratefully acknowledge financial support by the Deutsche 
Forschungsgemeinschaft and partial funding for their stay in Golden by the {\em
Particulate Science \& Technology Group} of the Colorado School of Mines. JLM
and MN are supported in part by NASA through contract number NAG3-1970. We also
would like to thank Susan McCaffery for a critical reading of the manuscript. A
generous grant of computer time on the Cray T3E at the Forschungszentrum
J\"ulich made the numerical simulations possible.

\bibliographystyle{prsty}
\bibliography{habil}

\ifx\grdraft\undefined
\newpage
\newpage
\begin{table}[htb]
  \begin{center}
    \begin{tabular}{l||c|c|c|c|c|c}
    $g$ [m/s$^2$]       & 1.62 & 3.73 & 9.81 & 13.6 & 25.1 &  274 \\ \hline
    $\Theta$ [$^\circ$] & 48.2 & 41.4 & 35.0 & 33.3 & 30.2 & 18.4 \\ \hline
    $\zeta/R$           &0.291 &0.274 &0.277 & 0.269& 0.283& 0.260
    \end{tabular}
  \end{center}
  \caption{Angle of repose, $\Theta$, in the drum middle and dimensionless 
           characteristic length, $\zeta/R$, as function of gravity, $g$ 
           (simulation).}
  \label{tab: gravity}
\end{table}

\newpage
\subsection*{Figure Captions}
\begin{description}
\item[\bf Figure~\ref{fig: sketch}] (a) Flat surface for low rotation speeds, 
      (b) deformed surface for medium rotation speeds with two straight lines 
      added as approximation and (c) fully developed S-shaped surface for 
      higher rotation speeds.
\item[\bf Figure~\ref{fig: mustard_all}] Experimentally measured starting
      ($\circ$) and stopping ($\star$) angle and dynamic angle of repose 
      ($\bullet$) for mustard seeds.
\item[\bf Figure~\ref{fig: mustard}] Dynamic angle of repose for black
      ($\bullet$) and yellow ($\circ$) mustard seeds.
\item[\bf Figure~\ref{fig: glass}] Dynamic angle of repose for small ($\bullet$)
      and large ($\circ$) glass beads.
\item[\bf Figure~\ref{fig: mri}] Comparison of dynamic angle of repose for large
      mustard seeds taken from MRI ($\circ$) and non-MRI ($\star$) measurements.
\item[\bf Figure~\ref{fig: fit}] Comparison of dynamic angle of repose for large
      mustard seeds taken from MRI ($\circ$), numerical simulation ($\bullet$) 
      and the theory of Zik et al.~\cite{zik94} (---).
\item[\bf Figure~\ref{fig: profile}] Profile of the dynamic angle of repose
      along the rotation axis for 2.5 mm spheres: ($\bullet$) simulation, (---) 
      fit.
\item[\bf Figure~\ref{fig: angle}] Dynamic angle of repose as function of sphere
      diameter for $\Omega=20$ rpm (simulation); ($\star$) end cap, ($\bullet$) 
      drum middle, (---) arcus-tangent fit.
\item[\bf Figure~\ref{fig: range}] Dimensionless range of boundary effect for
      spheres with different diameter; ($\star$) $R=3.5$ cm, ($\circ$) $R=7$ 
      cm, ($\bullet$) $R=10.5$ cm, dotted line shows value $\zeta=0.28 \, R$ 
      for radius-independent regime (simulation).
\item[\bf Figure~\ref{fig: range3}] Range of boundary effect measured in sphere
      diameters for spheres with different diameter; ($\star$) $R=3.5$ cm, 
      ($\circ$) $R=7$ cm, ($\bullet$) $R=10.5$ cm, (---) exponential fit 
      (simulation).
\end{description}

\clearpage

\begin{figure}[htb]
  \begin{center}
    \epsfig{file=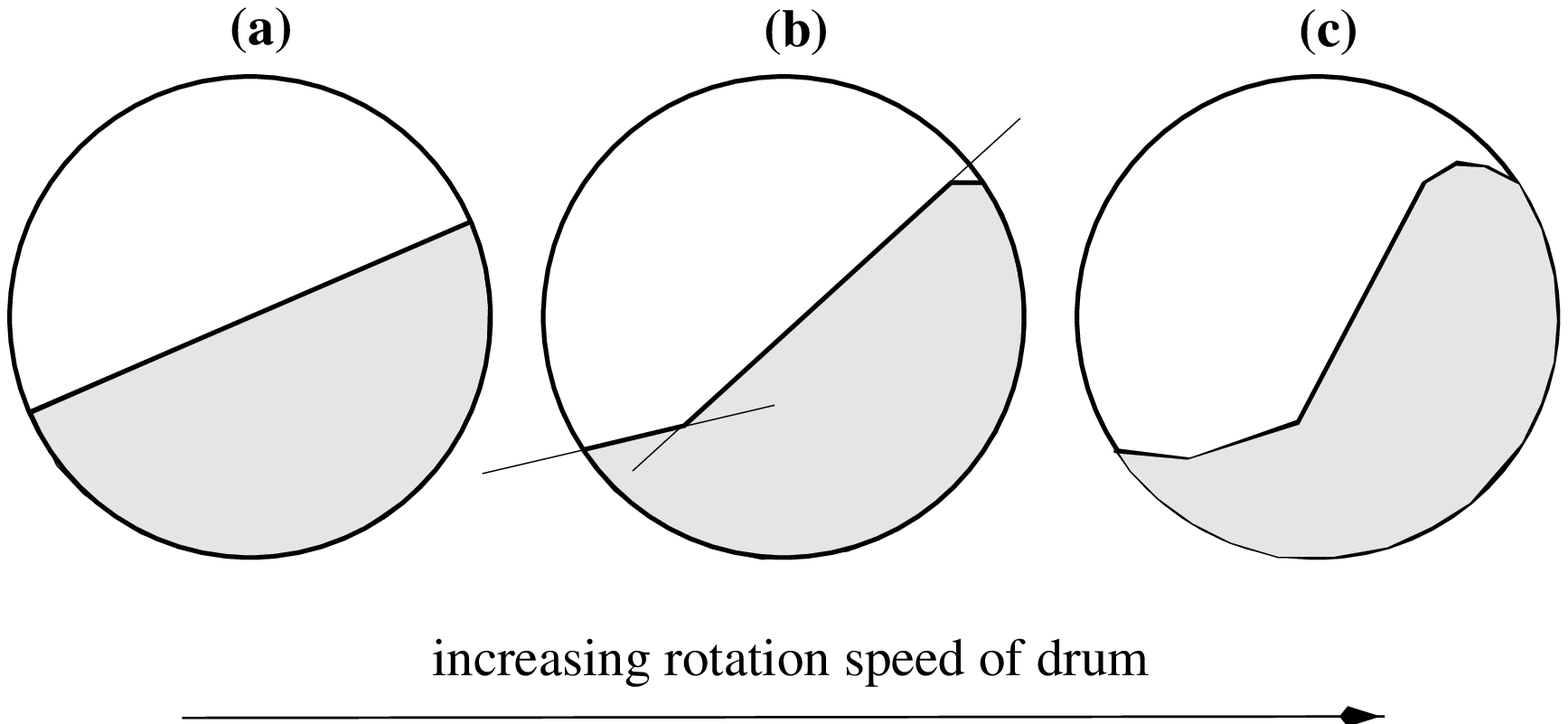,width=0.5\textwidth}
  \end{center}
  \caption{}
  \label{fig: sketch}
\end{figure}

\begin{figure}[htb]
  \begin{center}
\setlength{\unitlength}{0.240900pt}
\begin{picture}(1049,809)(0,0)
\thicklines \path(220,113)(240,113)
\thicklines \path(985,113)(965,113)
\put(198,113){\makebox(0,0)[r]{26}}
\thicklines \path(220,281)(240,281)
\thicklines \path(985,281)(965,281)
\put(198,281){\makebox(0,0)[r]{30}}
\thicklines \path(220,450)(240,450)
\thicklines \path(985,450)(965,450)
\put(198,450){\makebox(0,0)[r]{34}}
\thicklines \path(220,618)(240,618)
\thicklines \path(985,618)(965,618)
\put(198,618){\makebox(0,0)[r]{38}}
\thicklines \path(220,786)(240,786)
\thicklines \path(985,786)(965,786)
\put(198,786){\makebox(0,0)[r]{42}}
\thicklines \path(220,113)(220,133)
\thicklines \path(220,786)(220,766)
\put(220,68){\makebox(0,0){0}}
\thicklines \path(411,113)(411,133)
\thicklines \path(411,786)(411,766)
\put(411,68){\makebox(0,0){5}}
\thicklines \path(603,113)(603,133)
\thicklines \path(603,786)(603,766)
\put(603,68){\makebox(0,0){10}}
\thicklines \path(794,113)(794,133)
\thicklines \path(794,786)(794,766)
\put(794,68){\makebox(0,0){15}}
\thicklines \path(985,113)(985,133)
\thicklines \path(985,786)(985,766)
\put(985,68){\makebox(0,0){20}}
\thicklines \path(220,113)(985,113)(985,786)(220,786)(220,113)
\put(45,449){\makebox(0,0)[l]{\shortstack{$\langle\Theta\rangle$}}}
\put(602,0){\makebox(0,0){$\Omega$ [rpm]}}
\put(243,481){\circle{18}}
\put(252,505){\circle{18}}
\put(263,534){\circle{18}}
\put(266,500){\circle{18}}
\put(277,555){\circle{18}}
\put(284,492){\circle{18}}
\put(293,502){\circle{18}}
\put(307,475){\circle{18}}
\put(319,502){\circle{18}}
\put(333,492){\circle{18}}
\put(346,550){\circle{18}}
\thinlines \path(243,446)(243,516)
\thinlines \path(233,446)(253,446)
\thinlines \path(233,516)(253,516)
\thinlines \path(252,486)(252,525)
\thinlines \path(242,486)(262,486)
\thinlines \path(242,525)(262,525)
\thinlines \path(263,534)(263,534)
\thinlines \path(253,534)(273,534)
\thinlines \path(253,534)(273,534)
\thinlines \path(266,468)(266,532)
\thinlines \path(256,468)(276,468)
\thinlines \path(256,532)(276,532)
\thinlines \path(277,534)(277,576)
\thinlines \path(267,534)(287,534)
\thinlines \path(267,576)(287,576)
\thinlines \path(284,492)(284,492)
\thinlines \path(274,492)(294,492)
\thinlines \path(274,492)(294,492)
\thinlines \path(293,484)(293,520)
\thinlines \path(283,484)(303,484)
\thinlines \path(283,520)(303,520)
\thinlines \path(307,454)(307,495)
\thinlines \path(297,454)(317,454)
\thinlines \path(297,495)(317,495)
\thinlines \path(319,484)(319,520)
\thinlines \path(309,484)(329,484)
\thinlines \path(309,520)(329,520)
\thinlines \path(333,492)(333,492)
\thinlines \path(323,492)(343,492)
\thinlines \path(323,492)(343,492)
\thinlines \path(346,530)(346,571)
\thinlines \path(336,530)(356,530)
\thinlines \path(336,571)(356,571)
\put(243,239){\makebox(0,0){$\star$}}
\put(252,379){\makebox(0,0){$\star$}}
\put(263,323){\makebox(0,0){$\star$}}
\put(266,292){\makebox(0,0){$\star$}}
\put(277,334){\makebox(0,0){$\star$}}
\put(284,309){\makebox(0,0){$\star$}}
\put(293,334){\makebox(0,0){$\star$}}
\put(307,340){\makebox(0,0){$\star$}}
\put(319,334){\makebox(0,0){$\star$}}
\put(333,323){\makebox(0,0){$\star$}}
\put(346,365){\makebox(0,0){$\star$}}
\thinlines \path(243,209)(243,269)
\thinlines \path(233,209)(253,209)
\thinlines \path(233,269)(253,269)
\thinlines \path(252,359)(252,399)
\thinlines \path(242,359)(262,359)
\thinlines \path(242,399)(262,399)
\thinlines \path(263,323)(263,323)
\thinlines \path(253,323)(273,323)
\thinlines \path(253,323)(273,323)
\thinlines \path(266,274)(266,310)
\thinlines \path(256,274)(276,274)
\thinlines \path(256,310)(276,310)
\thinlines \path(277,299)(277,369)
\thinlines \path(267,299)(287,299)
\thinlines \path(267,369)(287,369)
\thinlines \path(284,290)(284,329)
\thinlines \path(274,290)(294,290)
\thinlines \path(274,329)(294,329)
\thinlines \path(293,299)(293,369)
\thinlines \path(283,299)(303,299)
\thinlines \path(283,369)(303,369)
\thinlines \path(307,320)(307,361)
\thinlines \path(297,320)(317,320)
\thinlines \path(297,361)(317,361)
\thinlines \path(319,316)(319,352)
\thinlines \path(309,316)(329,316)
\thinlines \path(309,352)(329,352)
\thinlines \path(333,323)(333,323)
\thinlines \path(323,323)(343,323)
\thinlines \path(323,323)(343,323)
\thinlines \path(346,339)(346,392)
\thinlines \path(336,339)(356,339)
\thinlines \path(336,392)(356,392)
\put(349,460){\circle*{12}}
\put(356,466){\circle*{12}}
\put(365,475){\circle*{12}}
\put(367,450){\circle*{12}}
\put(415,492){\circle*{12}}
\put(495,534){\circle*{12}}
\put(633,576){\circle*{12}}
\put(748,618){\circle*{12}}
\put(909,660){\circle*{12}}
\thinlines \path(349,442)(349,478)
\thinlines \path(339,442)(359,442)
\thinlines \path(339,478)(359,478)
\thinlines \path(356,433)(356,500)
\thinlines \path(346,433)(366,433)
\thinlines \path(346,500)(366,500)
\thinlines \path(365,454)(365,495)
\thinlines \path(355,454)(375,454)
\thinlines \path(355,495)(375,495)
\thinlines \path(367,420)(367,479)
\thinlines \path(357,420)(377,420)
\thinlines \path(357,479)(377,479)
\thinlines \path(415,492)(415,492)
\thinlines \path(405,492)(425,492)
\thinlines \path(405,492)(425,492)
\thinlines \path(495,534)(495,534)
\thinlines \path(485,534)(505,534)
\thinlines \path(485,534)(505,534)
\thinlines \path(633,576)(633,576)
\thinlines \path(623,576)(643,576)
\thinlines \path(623,576)(643,576)
\thinlines \path(748,618)(748,618)
\thinlines \path(738,618)(758,618)
\thinlines \path(738,618)(758,618)
\thinlines \path(909,660)(909,660)
\thinlines \path(899,660)(919,660)
\thinlines \path(899,660)(919,660)
\thicklines \path(349,460)(349,460)(356,466)(365,475)(367,450)(415,492)(495,534)(633,576)(748,618)(909,660)
\end{picture}
  \end{center}
  \caption{}
  \label{fig: mustard_all}
\end{figure}

\begin{figure}[htb]
  \begin{center}
\setlength{\unitlength}{0.240900pt}
\begin{picture}(1049,809)(0,0)
\thicklines \path(220,113)(240,113)
\thicklines \path(985,113)(965,113)
\put(198,113){\makebox(0,0)[r]{32}}
\thicklines \path(220,281)(240,281)
\thicklines \path(985,281)(965,281)
\put(198,281){\makebox(0,0)[r]{36}}
\thicklines \path(220,450)(240,450)
\thicklines \path(985,450)(965,450)
\put(198,450){\makebox(0,0)[r]{40}}
\thicklines \path(220,618)(240,618)
\thicklines \path(985,618)(965,618)
\put(198,618){\makebox(0,0)[r]{44}}
\thicklines \path(220,786)(240,786)
\thicklines \path(985,786)(965,786)
\put(198,786){\makebox(0,0)[r]{48}}
\thicklines \path(220,113)(220,133)
\thicklines \path(220,786)(220,766)
\put(220,68){\makebox(0,0){0}}
\thicklines \path(390,113)(390,133)
\thicklines \path(390,786)(390,766)
\put(390,68){\makebox(0,0){10}}
\thicklines \path(560,113)(560,133)
\thicklines \path(560,786)(560,766)
\put(560,68){\makebox(0,0){20}}
\thicklines \path(730,113)(730,133)
\thicklines \path(730,786)(730,766)
\put(730,68){\makebox(0,0){30}}
\thicklines \path(900,113)(900,133)
\thicklines \path(900,786)(900,766)
\put(900,68){\makebox(0,0){40}}
\thicklines \path(220,113)(985,113)(985,786)(220,786)(220,113)
\put(45,449){\makebox(0,0)[l]{\shortstack{$\langle\Theta\rangle$}}}
\put(602,0){\makebox(0,0){$\Omega$ [rpm]}}
\put(475,702){\makebox(0,0)[r]{2.5 mm}}
\put(519,702){\circle{18}}
\put(305,244){\circle{18}}
\put(327,251){\circle{18}}
\put(348,259){\circle{18}}
\put(370,287){\circle{18}}
\put(390,289){\circle{18}}
\put(433,301){\circle{18}}
\put(475,360){\circle{18}}
\put(518,413){\circle{18}}
\put(560,431){\circle{18}}
\put(645,503){\circle{18}}
\put(730,560){\circle{18}}
\put(900,660){\circle{18}}
\thinlines \path(497,702)(563,702)
\thinlines \path(497,712)(497,692)
\thinlines \path(563,712)(563,692)
\thinlines \path(305,234)(305,255)
\thinlines \path(295,234)(315,234)
\thinlines \path(295,255)(315,255)
\thinlines \path(327,237)(327,265)
\thinlines \path(317,237)(337,237)
\thinlines \path(317,265)(337,265)
\thinlines \path(348,248)(348,270)
\thinlines \path(338,248)(358,248)
\thinlines \path(338,270)(358,270)
\thinlines \path(370,276)(370,297)
\thinlines \path(360,276)(380,276)
\thinlines \path(360,297)(380,297)
\thinlines \path(390,279)(390,299)
\thinlines \path(380,279)(400,279)
\thinlines \path(380,299)(400,299)
\thinlines \path(433,292)(433,311)
\thinlines \path(423,292)(443,292)
\thinlines \path(423,311)(443,311)
\thinlines \path(475,347)(475,373)
\thinlines \path(465,347)(485,347)
\thinlines \path(465,373)(485,373)
\thinlines \path(518,407)(518,418)
\thinlines \path(508,407)(528,407)
\thinlines \path(508,418)(528,418)
\thinlines \path(560,422)(560,439)
\thinlines \path(550,422)(570,422)
\thinlines \path(550,439)(570,439)
\thinlines \path(645,492)(645,514)
\thinlines \path(635,492)(655,492)
\thinlines \path(635,514)(655,514)
\thinlines \path(730,537)(730,583)
\thinlines \path(720,537)(740,537)
\thinlines \path(720,583)(740,583)
\thinlines \path(900,648)(900,672)
\thinlines \path(890,648)(910,648)
\thinlines \path(890,672)(910,672)
\put(475,657){\makebox(0,0)[r]{1.7 mm}}
\put(519,657){\circle*{12}}
\put(285,166){\circle*{12}}
\put(305,160){\circle*{12}}
\put(327,176){\circle*{12}}
\put(348,166){\circle*{12}}
\put(370,187){\circle*{12}}
\put(390,192){\circle*{12}}
\put(433,262){\circle*{12}}
\put(475,322){\circle*{12}}
\put(518,357){\circle*{12}}
\put(560,420){\circle*{12}}
\put(603,441){\circle*{12}}
\put(645,502){\circle*{12}}
\put(730,565){\circle*{12}}
\put(815,660){\circle*{12}}
\put(900,704){\circle*{12}}
\thinlines \path(497,657)(563,657)
\thinlines \path(497,667)(497,647)
\thinlines \path(563,667)(563,647)
\thinlines \path(285,160)(285,171)
\thinlines \path(275,160)(295,160)
\thinlines \path(275,171)(295,171)
\thinlines \path(305,155)(305,166)
\thinlines \path(295,155)(315,155)
\thinlines \path(295,166)(315,166)
\thinlines \path(327,164)(327,188)
\thinlines \path(317,164)(337,164)
\thinlines \path(317,188)(337,188)
\thinlines \path(348,156)(348,176)
\thinlines \path(338,156)(358,156)
\thinlines \path(338,176)(358,176)
\thinlines \path(370,181)(370,192)
\thinlines \path(360,181)(380,181)
\thinlines \path(360,192)(380,192)
\thinlines \path(390,182)(390,202)
\thinlines \path(380,182)(400,182)
\thinlines \path(380,202)(400,202)
\thinlines \path(433,255)(433,269)
\thinlines \path(423,255)(443,255)
\thinlines \path(423,269)(443,269)
\thinlines \path(475,314)(475,330)
\thinlines \path(465,314)(485,314)
\thinlines \path(465,330)(485,330)
\thinlines \path(518,347)(518,367)
\thinlines \path(508,347)(528,347)
\thinlines \path(508,367)(528,367)
\thinlines \path(560,415)(560,425)
\thinlines \path(550,415)(570,415)
\thinlines \path(550,425)(570,425)
\thinlines \path(603,429)(603,453)
\thinlines \path(593,429)(613,429)
\thinlines \path(593,453)(613,453)
\thinlines \path(645,496)(645,508)
\thinlines \path(635,496)(655,496)
\thinlines \path(635,508)(655,508)
\thinlines \path(730,559)(730,571)
\thinlines \path(720,559)(740,559)
\thinlines \path(720,571)(740,571)
\thinlines \path(815,649)(815,670)
\thinlines \path(805,649)(825,649)
\thinlines \path(805,670)(825,670)
\thinlines \path(900,697)(900,711)
\thinlines \path(890,697)(910,697)
\thinlines \path(890,711)(910,711)
\end{picture}
  \end{center}
  \caption{}
  \label{fig: mustard}
\end{figure}

\begin{figure}[htb]
  \begin{center}
\setlength{\unitlength}{0.240900pt}
\begin{picture}(1049,809)(0,0)
\thicklines \path(220,113)(240,113)
\thicklines \path(985,113)(965,113)
\put(198,113){\makebox(0,0)[r]{28}}
\thicklines \path(220,281)(240,281)
\thicklines \path(985,281)(965,281)
\put(198,281){\makebox(0,0)[r]{32}}
\thicklines \path(220,450)(240,450)
\thicklines \path(985,450)(965,450)
\put(198,450){\makebox(0,0)[r]{36}}
\thicklines \path(220,618)(240,618)
\thicklines \path(985,618)(965,618)
\put(198,618){\makebox(0,0)[r]{40}}
\thicklines \path(220,786)(240,786)
\thicklines \path(985,786)(965,786)
\put(198,786){\makebox(0,0)[r]{44}}
\thicklines \path(220,113)(220,133)
\thicklines \path(220,786)(220,766)
\put(220,68){\makebox(0,0){0}}
\thicklines \path(390,113)(390,133)
\thicklines \path(390,786)(390,766)
\put(390,68){\makebox(0,0){10}}
\thicklines \path(560,113)(560,133)
\thicklines \path(560,786)(560,766)
\put(560,68){\makebox(0,0){20}}
\thicklines \path(730,113)(730,133)
\thicklines \path(730,786)(730,766)
\put(730,68){\makebox(0,0){30}}
\thicklines \path(900,113)(900,133)
\thicklines \path(900,786)(900,766)
\put(900,68){\makebox(0,0){40}}
\thicklines \path(220,113)(985,113)(985,786)(220,786)(220,113)
\put(45,449){\makebox(0,0)[l]{\shortstack{$\langle\Theta\rangle$}}}
\put(602,0){\makebox(0,0){$\Omega$ [rpm]}}
\put(475,702){\makebox(0,0)[r]{1.5 mm}}
\put(519,702){\circle*{12}}
\put(285,171){\circle*{12}}
\put(305,181){\circle*{12}}
\put(327,196){\circle*{12}}
\put(348,209){\circle*{12}}
\put(370,227){\circle*{12}}
\put(390,237){\circle*{12}}
\put(433,257){\circle*{12}}
\put(475,269){\circle*{12}}
\put(518,315){\circle*{12}}
\put(560,357){\circle*{12}}
\put(645,444){\circle*{12}}
\put(730,518){\circle*{12}}
\put(815,618){\circle*{12}}
\put(900,691){\circle*{12}}
\thinlines \path(497,702)(563,702)
\thinlines \path(497,712)(497,692)
\thinlines \path(563,712)(563,692)
\thinlines \path(285,162)(285,180)
\thinlines \path(275,162)(295,162)
\thinlines \path(275,180)(295,180)
\thinlines \path(305,173)(305,190)
\thinlines \path(295,173)(315,173)
\thinlines \path(295,190)(315,190)
\thinlines \path(327,192)(327,200)
\thinlines \path(317,192)(337,192)
\thinlines \path(317,200)(337,200)
\thinlines \path(348,199)(348,218)
\thinlines \path(338,199)(358,199)
\thinlines \path(338,218)(358,218)
\thinlines \path(370,220)(370,233)
\thinlines \path(360,220)(380,220)
\thinlines \path(360,233)(380,233)
\thinlines \path(390,228)(390,246)
\thinlines \path(380,228)(400,228)
\thinlines \path(380,246)(400,246)
\thinlines \path(433,249)(433,265)
\thinlines \path(423,249)(443,249)
\thinlines \path(423,265)(443,265)
\thinlines \path(475,262)(475,275)
\thinlines \path(465,262)(485,262)
\thinlines \path(465,275)(485,275)
\thinlines \path(518,305)(518,325)
\thinlines \path(508,305)(528,305)
\thinlines \path(508,325)(528,325)
\thinlines \path(560,351)(560,363)
\thinlines \path(550,351)(570,351)
\thinlines \path(550,363)(570,363)
\thinlines \path(645,429)(645,460)
\thinlines \path(635,429)(655,429)
\thinlines \path(635,460)(655,460)
\thinlines \path(730,509)(730,527)
\thinlines \path(720,509)(740,509)
\thinlines \path(720,527)(740,527)
\thinlines \path(815,605)(815,631)
\thinlines \path(805,605)(825,605)
\thinlines \path(805,631)(825,631)
\thinlines \path(900,686)(900,697)
\thinlines \path(890,686)(910,686)
\thinlines \path(890,697)(910,697)
\put(475,657){\makebox(0,0)[r]{3.0 mm}}
\put(519,657){\circle{18}}
\put(305,187){\circle{18}}
\put(327,192){\circle{18}}
\put(348,199){\circle{18}}
\put(370,222){\circle{18}}
\put(390,244){\circle{18}}
\put(433,250){\circle{18}}
\put(475,289){\circle{18}}
\put(518,297){\circle{18}}
\put(560,310){\circle{18}}
\put(603,318){\circle{18}}
\put(645,355){\circle{18}}
\put(730,413){\circle{18}}
\put(815,492){\circle{18}}
\put(900,544){\circle{18}}
\thinlines \path(497,657)(563,657)
\thinlines \path(497,667)(497,647)
\thinlines \path(563,667)(563,647)
\thinlines \path(305,173)(305,200)
\thinlines \path(295,173)(315,173)
\thinlines \path(295,200)(315,200)
\thinlines \path(327,179)(327,205)
\thinlines \path(317,179)(337,179)
\thinlines \path(317,205)(337,205)
\thinlines \path(348,190)(348,208)
\thinlines \path(338,190)(358,190)
\thinlines \path(338,208)(358,208)
\thinlines \path(370,209)(370,235)
\thinlines \path(360,209)(380,209)
\thinlines \path(360,235)(380,235)
\thinlines \path(390,239)(390,250)
\thinlines \path(380,239)(400,239)
\thinlines \path(380,250)(400,250)
\thinlines \path(433,230)(433,270)
\thinlines \path(423,230)(443,230)
\thinlines \path(423,270)(443,270)
\thinlines \path(475,277)(475,300)
\thinlines \path(465,277)(485,277)
\thinlines \path(465,300)(485,300)
\thinlines \path(518,287)(518,307)
\thinlines \path(508,287)(528,287)
\thinlines \path(508,307)(528,307)
\thinlines \path(560,305)(560,315)
\thinlines \path(550,305)(570,305)
\thinlines \path(550,315)(570,315)
\thinlines \path(603,308)(603,328)
\thinlines \path(593,308)(613,308)
\thinlines \path(593,328)(613,328)
\thinlines \path(645,327)(645,383)
\thinlines \path(635,327)(655,327)
\thinlines \path(635,383)(655,383)
\thinlines \path(730,407)(730,418)
\thinlines \path(720,407)(740,407)
\thinlines \path(720,418)(740,418)
\thinlines \path(815,481)(815,502)
\thinlines \path(805,481)(825,481)
\thinlines \path(805,502)(825,502)
\thinlines \path(900,534)(900,555)
\thinlines \path(890,534)(910,534)
\thinlines \path(890,555)(910,555)
\end{picture}
  \end{center}
  \caption{}
  \label{fig: glass}
\end{figure}

\begin{figure}[htb]
  \begin{center}
\setlength{\unitlength}{0.240900pt}
\begin{picture}(1049,809)(0,0)
\thicklines \path(220,113)(240,113)
\thicklines \path(985,113)(965,113)
\put(198,113){\makebox(0,0)[r]{28}}
\thicklines \path(220,337)(240,337)
\thicklines \path(985,337)(965,337)
\put(198,337){\makebox(0,0)[r]{32}}
\thicklines \path(220,562)(240,562)
\thicklines \path(985,562)(965,562)
\put(198,562){\makebox(0,0)[r]{36}}
\thicklines \path(220,786)(240,786)
\thicklines \path(985,786)(965,786)
\put(198,786){\makebox(0,0)[r]{40}}
\thicklines \path(220,113)(220,133)
\thicklines \path(220,786)(220,766)
\put(220,68){\makebox(0,0){0}}
\thicklines \path(439,113)(439,133)
\thicklines \path(439,786)(439,766)
\put(439,68){\makebox(0,0){10}}
\thicklines \path(657,113)(657,133)
\thicklines \path(657,786)(657,766)
\put(657,68){\makebox(0,0){20}}
\thicklines \path(876,113)(876,133)
\thicklines \path(876,786)(876,766)
\put(876,68){\makebox(0,0){30}}
\thicklines \path(220,113)(985,113)(985,786)(220,786)(220,113)
\put(45,449){\makebox(0,0)[l]{\shortstack{$\langle\Theta\rangle$}}}
\put(602,0){\makebox(0,0){$\Omega$ [rpm]}}
\put(821,281){\makebox(0,0)[r]{non-MRI}}
\put(865,281){\makebox(0,0){$\star$}}
\put(294,466){\makebox(0,0){$\star$}}
\put(333,506){\makebox(0,0){$\star$}}
\put(376,561){\makebox(0,0){$\star$}}
\put(456,618){\makebox(0,0){$\star$}}
\put(517,673){\makebox(0,0){$\star$}}
\put(584,729){\makebox(0,0){$\star$}}
\put(821,236){\makebox(0,0)[r]{MRI}}
\put(865,236){\circle{18}}
\put(322,225){\circle{18}}
\put(404,281){\circle{18}}
\put(469,336){\circle{18}}
\put(522,393){\circle{18}}
\put(600,450){\circle{18}}
\put(666,506){\circle{18}}
\put(745,561){\circle{18}}
\put(810,616){\circle{18}}
\put(889,673){\circle{18}}
\end{picture}
  \end{center}
  \caption{}
   \label{fig: mri}
\end{figure}

\begin{figure}[htb]
  \begin{center}
\setlength{\unitlength}{0.240900pt}
\begin{picture}(1049,809)(0,0)
\thicklines \path(220,113)(240,113)
\thicklines \path(985,113)(965,113)
\put(198,113){\makebox(0,0)[r]{24}}
\thicklines \path(220,281)(240,281)
\thicklines \path(985,281)(965,281)
\put(198,281){\makebox(0,0)[r]{28}}
\thicklines \path(220,450)(240,450)
\thicklines \path(985,450)(965,450)
\put(198,450){\makebox(0,0)[r]{32}}
\thicklines \path(220,618)(240,618)
\thicklines \path(985,618)(965,618)
\put(198,618){\makebox(0,0)[r]{36}}
\thicklines \path(220,786)(240,786)
\thicklines \path(985,786)(965,786)
\put(198,786){\makebox(0,0)[r]{40}}
\thicklines \path(220,113)(220,133)
\thicklines \path(220,786)(220,766)
\put(220,68){\makebox(0,0){0}}
\thicklines \path(439,113)(439,133)
\thicklines \path(439,786)(439,766)
\put(439,68){\makebox(0,0){10}}
\thicklines \path(657,113)(657,133)
\thicklines \path(657,786)(657,766)
\put(657,68){\makebox(0,0){20}}
\thicklines \path(876,113)(876,133)
\thicklines \path(876,786)(876,766)
\put(876,68){\makebox(0,0){30}}
\thicklines \path(220,113)(985,113)(985,786)(220,786)(220,113)
\put(45,449){\makebox(0,0)[l]{\shortstack{$\langle\Theta\rangle$}}}
\put(602,0){\makebox(0,0){$\Omega$ [rpm]}}
\put(854,281){\makebox(0,0)[r]{simulation}}
\put(898,281){\circle*{12}}
\put(293,271){( \circle*{12} )}
\put(377,357){\circle*{12}}
\put(429,424){\circle*{12}}
\put(533,511){\circle*{12}}
\put(637,576){\circle*{12}}
\put(742,630){\circle*{12}}
\put(846,676){\circle*{12}}
\put(898,696){\circle*{12}}
\thinlines \path(876,281)(942,281)
\thinlines \path(876,291)(876,271)
\thinlines \path(942,291)(942,271)
\thinlines \path(324,266)(324,276)
\thinlines \path(314,266)(334,266)
\thinlines \path(314,276)(334,276)
\thinlines \path(377,354)(377,360)
\thinlines \path(367,354)(387,354)
\thinlines \path(367,360)(387,360)
\thinlines \path(429,420)(429,428)
\thinlines \path(419,420)(439,420)
\thinlines \path(419,428)(439,428)
\thinlines \path(533,508)(533,514)
\thinlines \path(523,508)(543,508)
\thinlines \path(523,514)(543,514)
\thinlines \path(637,573)(637,580)
\thinlines \path(627,573)(647,573)
\thinlines \path(627,580)(647,580)
\thinlines \path(742,627)(742,634)
\thinlines \path(732,627)(752,627)
\thinlines \path(732,634)(752,634)
\thinlines \path(846,672)(846,680)
\thinlines \path(836,672)(856,672)
\thinlines \path(836,680)(856,680)
\thinlines \path(898,692)(898,699)
\thinlines \path(888,692)(908,692)
\thinlines \path(888,699)(908,699)
\put(854,236){\makebox(0,0)[r]{experiment}}
\put(898,236){\circle{18}}
\put(322,365){\circle{18}}
\put(404,407){\circle{18}}
\put(469,449){\circle{18}}
\put(522,491){\circle{18}}
\put(600,534){\circle{18}}
\put(666,576){\circle{18}}
\put(745,617){\circle{18}}
\put(810,658){\circle{18}}
\put(889,701){\circle{18}}
\put(854,191){\makebox(0,0)[r]{theory}}
\thinlines \path(876,191)(942,191)
\thinlines \path(325,344)(325,344)(431,427)(536,499)(642,565)(747,625)(852,680)(958,732)
\put(854,146){\makebox(0,0)[r]{quadratic fit}}
\dottedline{12}(876,146)(942,146)
\dottedline{12}(313,371)(320,373)(328,375)(336,378)(344,380)(351,383)(359,386)(367,390)(375,393)(382,397)(390,400)(398,404)(405,408)(413,413)(421,417)(429,422)(436,426)(444,431)(452,437)(460,442)(467,447)(475,453)(483,459)(490,465)(498,471)(506,478)(514,484)(521,491)(529,498)(537,505)(545,512)(552,519)(560,527)(568,535)(575,543)(583,551)(591,559)(599,568)
\dottedline{12}(599,568)(606,576)(614,585)(622,594)(630,603)(637,612)(645,622)(653,632)(660,641)(668,652)(676,662)(684,672)(691,683)(699,693)(707,704)(715,715)(722,727)(730,738)(738,750)(745,761)(753,773)(761,786)(761,786)
\end{picture}
  \end{center}
  \caption[]{}
   \label{fig: fit}
\end{figure}

\begin{figure}[htb]
  \begin{center}
    \input bound_6
  \end{center}
  \caption{}
   \label{fig: profile}
\end{figure}

\begin{figure}[htb]
  \begin{center}
\setlength{\unitlength}{0.240900pt}
\begin{picture}(1049,809)(0,0)
\thicklines \path(220,113)(240,113)
\thicklines \path(985,113)(965,113)
\put(198,113){\makebox(0,0)[r]{26}}
\thicklines \path(220,248)(240,248)
\thicklines \path(985,248)(965,248)
\put(198,248){\makebox(0,0)[r]{30}}
\thicklines \path(220,382)(240,382)
\thicklines \path(985,382)(965,382)
\put(198,382){\makebox(0,0)[r]{34}}
\thicklines \path(220,517)(240,517)
\thicklines \path(985,517)(965,517)
\put(198,517){\makebox(0,0)[r]{38}}
\thicklines \path(220,651)(240,651)
\thicklines \path(985,651)(965,651)
\put(198,651){\makebox(0,0)[r]{42}}
\thicklines \path(220,786)(240,786)
\thicklines \path(985,786)(965,786)
\put(198,786){\makebox(0,0)[r]{46}}
\thicklines \path(220,113)(220,133)
\thicklines \path(220,786)(220,766)
\put(220,68){\makebox(0,0){0}}
\thicklines \path(411,113)(411,133)
\thicklines \path(411,786)(411,766)
\put(411,68){\makebox(0,0){2}}
\thicklines \path(603,113)(603,133)
\thicklines \path(603,786)(603,766)
\put(603,68){\makebox(0,0){4}}
\thicklines \path(794,113)(794,133)
\thicklines \path(794,786)(794,766)
\put(794,68){\makebox(0,0){6}}
\thicklines \path(985,113)(985,133)
\thicklines \path(985,786)(985,766)
\put(985,68){\makebox(0,0){8}}
\thicklines \path(220,113)(985,113)(985,786)(220,786)(220,113)
\put(45,449){\makebox(0,0)[l]{\shortstack{$\langle\Theta\rangle$}}}
\put(602,23){\makebox(0,0){$d$ [mm]}}
\put(842,248){\makebox(0,0)[r]{end cap}}
\put(886,248){\makebox(0,0){$\star$}}
\put(889,749){\makebox(0,0){$\star$}}
\put(794,715){\makebox(0,0){$\star$}}
\put(698,688){\makebox(0,0){$\star$}}
\put(603,655){\makebox(0,0){$\star$}}
\put(555,624){\makebox(0,0){$\star$}}
\put(507,591){\makebox(0,0){$\star$}}
\put(459,571){\makebox(0,0){$\star$}}
\put(411,537){\makebox(0,0){$\star$}}
\put(363,490){\makebox(0,0){$\star$}}
\put(316,416){\makebox(0,0){$\star$}}
\thinlines \path(864,248)(930,248)
\thinlines \path(864,258)(864,238)
\thinlines \path(930,258)(930,238)
\thinlines \path(889,730)(889,767)
\thinlines \path(879,730)(899,730)
\thinlines \path(879,767)(899,767)
\thinlines \path(794,697)(794,734)
\thinlines \path(784,697)(804,697)
\thinlines \path(784,734)(804,734)
\thinlines \path(698,673)(698,704)
\thinlines \path(688,673)(708,673)
\thinlines \path(688,704)(708,704)
\thinlines \path(603,640)(603,670)
\thinlines \path(593,640)(613,640)
\thinlines \path(593,670)(613,670)
\thinlines \path(555,614)(555,635)
\thinlines \path(545,614)(565,614)
\thinlines \path(545,635)(565,635)
\thinlines \path(507,576)(507,606)
\thinlines \path(497,576)(517,576)
\thinlines \path(497,606)(517,606)
\thinlines \path(459,564)(459,577)
\thinlines \path(449,564)(469,564)
\thinlines \path(449,577)(469,577)
\thinlines \path(411,529)(411,545)
\thinlines \path(401,529)(421,529)
\thinlines \path(401,545)(421,545)
\thinlines \path(363,483)(363,497)
\thinlines \path(353,483)(373,483)
\thinlines \path(353,497)(373,497)
\thinlines \path(316,409)(316,423)
\thinlines \path(306,409)(326,409)
\thinlines \path(306,423)(326,423)
\put(842,203){\makebox(0,0)[r]{middle}}
\put(886,203){\circle*{12}}
\put(889,645){\circle*{12}}
\put(794,608){\circle*{12}}
\put(698,567){\circle*{12}}
\put(603,517){\circle*{12}}
\put(555,480){\circle*{12}}
\put(507,453){\circle*{12}}
\put(459,416){\circle*{12}}
\put(411,369){\circle*{12}}
\put(363,315){\circle*{12}}
\put(316,238){\circle*{12}}
\thinlines \path(864,203)(930,203)
\thinlines \path(864,213)(864,193)
\thinlines \path(930,213)(930,193)
\thinlines \path(889,635)(889,655)
\thinlines \path(879,635)(899,635)
\thinlines \path(879,655)(899,655)
\thinlines \path(794,594)(794,621)
\thinlines \path(784,594)(804,594)
\thinlines \path(784,621)(804,621)
\thinlines \path(698,557)(698,577)
\thinlines \path(688,557)(708,557)
\thinlines \path(688,577)(708,577)
\thinlines \path(603,507)(603,527)
\thinlines \path(593,507)(613,507)
\thinlines \path(593,527)(613,527)
\thinlines \path(555,473)(555,487)
\thinlines \path(545,473)(565,473)
\thinlines \path(545,487)(565,487)
\thinlines \path(507,439)(507,466)
\thinlines \path(497,439)(517,439)
\thinlines \path(497,466)(517,466)
\thinlines \path(459,412)(459,419)
\thinlines \path(449,412)(469,412)
\thinlines \path(449,419)(469,419)
\thinlines \path(411,362)(411,375)
\thinlines \path(401,362)(421,362)
\thinlines \path(401,375)(421,375)
\thinlines \path(363,308)(363,322)
\thinlines \path(353,308)(373,308)
\thinlines \path(353,322)(373,322)
\thinlines \path(316,231)(316,244)
\thinlines \path(306,231)(326,231)
\thinlines \path(306,244)(326,244)
\thinlines \path(228,113)(235,124)(243,136)(251,148)(259,160)(266,172)(274,184)(282,195)(290,207)(297,218)(305,229)(313,240)(320,251)(328,262)(336,272)(344,283)(351,293)(359,303)(367,313)(375,322)(382,332)(390,341)(398,350)(405,359)(413,367)(421,375)(429,384)(436,392)(444,399)(452,407)(460,414)(467,422)(475,429)(483,435)(490,442)(498,449)(506,455)(514,461)(521,467)(529,473)(537,479)(545,484)(552,490)(560,495)(568,500)(575,505)(583,510)(591,515)(599,520)(606,524)(614,528)
\thinlines \path(614,528)(622,533)(630,537)(637,541)(645,545)(653,549)(660,553)(668,557)(676,560)(684,564)(691,567)(699,571)(707,574)(715,577)(722,580)(730,583)(738,586)(745,589)(753,592)(761,595)(769,598)(776,601)(784,603)(792,606)(800,608)(807,611)(815,613)(823,616)(830,618)(838,620)(846,623)(854,625)(861,627)(869,629)(877,631)(885,633)(892,635)(900,637)(908,639)(915,641)(923,643)(931,644)(939,646)(946,648)(954,650)(962,651)(970,653)(977,655)(985,656)
\thinlines \path(220,274)(220,274)(228,287)(235,299)(243,312)(251,324)(259,336)(266,348)(274,360)(282,372)(290,384)(297,395)(305,406)(313,417)(320,428)(328,438)(336,448)(344,458)(351,467)(359,477)(367,486)(375,494)(382,503)(390,511)(398,519)(405,526)(413,534)(421,541)(429,548)(436,554)(444,561)(452,567)(460,573)(467,579)(475,585)(483,590)(490,595)(498,600)(506,605)(514,610)(521,615)(529,619)(537,623)(545,627)(552,632)(560,635)(568,639)(575,643)(583,646)(591,650)(599,653)
\thinlines \path(599,653)(606,656)(614,660)(622,663)(630,666)(637,669)(645,671)(653,674)(660,677)(668,679)(676,682)(684,684)(691,687)(699,689)(707,691)(715,693)(722,696)(730,698)(738,700)(745,702)(753,704)(761,706)(769,707)(776,709)(784,711)(792,713)(800,714)(807,716)(815,718)(823,719)(830,721)(838,722)(846,724)(854,725)(861,726)(869,728)(877,729)(885,730)(892,732)(900,733)(908,734)(915,735)(923,737)(931,738)(939,739)(946,740)(954,741)(962,742)(970,743)(977,744)(985,745)
\end{picture}
  \end{center}
  \caption{}
   \label{fig: angle}
\end{figure}

\begin{figure}[htb]
  \begin{center}
\setlength{\unitlength}{0.240900pt}
\begin{picture}(1500,900)(0,0)
\thicklines \path(199,179)(219,179)
\thicklines \path(952,179)(932,179)
\put(177,179){\makebox(0,0)[r]{0}}
\thicklines \path(199,292)(219,292)
\thicklines \path(952,292)(932,292)
\put(177,292){\makebox(0,0)[r]{}}
\thicklines \path(199,404)(219,404)
\thicklines \path(952,404)(932,404)
\put(177,404){\makebox(0,0)[r]{0.2}}
\thicklines \path(199,517)(219,517)
\thicklines \path(952,517)(932,517)
\put(177,517){\makebox(0,0)[r]{}}
\thicklines \path(199,630)(219,630)
\thicklines \path(952,630)(932,630)
\put(177,630){\makebox(0,0)[r]{0.4}}
\thicklines \path(199,742)(219,742)
\thicklines \path(952,742)(932,742)
\put(177,742){\makebox(0,0)[r]{}}
\thicklines \path(199,855)(219,855)
\thicklines \path(952,855)(932,855)
\put(177,855){\makebox(0,0)[r]{0.6}}
\thicklines \path(199,179)(199,199)
\thicklines \path(199,855)(199,835)
\put(199,134){\makebox(0,0){0}}
\thicklines \path(378,179)(378,199)
\thicklines \path(378,855)(378,835)
\put(378,134){\makebox(0,0){0.05}}
\thicklines \path(558,179)(558,199)
\thicklines \path(558,855)(558,835)
\put(558,134){\makebox(0,0){0.1}}
\thicklines \path(737,179)(737,199)
\thicklines \path(737,855)(737,835)
\put(737,134){\makebox(0,0){0.15}}
\thicklines \path(916,179)(916,199)
\thicklines \path(916,855)(916,835)
\put(916,134){\makebox(0,0){0.2}}
\thicklines \path(199,179)(952,179)(952,855)(199,855)(199,179)
\put(45,517){\makebox(0,0)[l]{\shortstack{$\zeta/R$}}}
\put(575,67){\makebox(0,0){$d/R$}}
\put(805,348){\makebox(0,0)[r]{$R=3.5$ cm}}
\thinlines \path(827,348)(935,348)
\thinlines \path(827,358)(827,338)
\thinlines \path(935,358)(935,338)
\thinlines \path(916,610)(916,791)
\thinlines \path(906,610)(926,610)
\thinlines \path(906,791)(926,791)
\thinlines \path(814,549)(814,659)
\thinlines \path(804,549)(824,549)
\thinlines \path(804,659)(824,659)
\thinlines \path(711,453)(711,549)
\thinlines \path(701,453)(721,453)
\thinlines \path(701,549)(721,549)
\thinlines \path(660,443)(660,507)
\thinlines \path(650,443)(670,443)
\thinlines \path(650,507)(670,507)
\thinlines \path(609,449)(609,527)
\thinlines \path(599,449)(619,449)
\thinlines \path(599,527)(619,527)
\thinlines \path(558,465)(558,517)
\thinlines \path(548,465)(568,465)
\thinlines \path(548,517)(568,517)
\thinlines \path(506,443)(506,520)
\thinlines \path(496,443)(516,443)
\thinlines \path(496,520)(516,520)
\thinlines \path(455,472)(455,511)
\thinlines \path(445,472)(465,472)
\thinlines \path(445,511)(465,511)
\thinlines \path(404,475)(404,514)
\thinlines \path(394,475)(414,475)
\thinlines \path(394,514)(414,514)
\thinlines \path(353,485)(353,517)
\thinlines \path(343,485)(363,485)
\thinlines \path(343,517)(363,517)
\thinlines \path(301,533)(301,559)
\thinlines \path(291,533)(311,533)
\thinlines \path(291,559)(311,559)
\put(916,700){\makebox(0,0){$\star$}}
\put(814,604){\makebox(0,0){$\star$}}
\put(711,501){\makebox(0,0){$\star$}}
\put(660,475){\makebox(0,0){$\star$}}
\put(609,488){\makebox(0,0){$\star$}}
\put(558,491){\makebox(0,0){$\star$}}
\put(506,482){\makebox(0,0){$\star$}}
\put(455,491){\makebox(0,0){$\star$}}
\put(404,494){\makebox(0,0){$\star$}}
\put(353,501){\makebox(0,0){$\star$}}
\put(301,546){\makebox(0,0){$\star$}}
\put(881,348){\makebox(0,0){$\star$}}
\put(805,303){\makebox(0,0)[r]{$7$ cm}}
\put(353,504){\circle{18}}
\put(404,512){\circle{18}}
\put(506,503){\circle{18}}
\put(609,519){\circle{18}}
\put(711,533){\circle{18}}
\put(814,628){\circle{18}}
\put(916,662){\circle{18}}
\put(881,303){\circle{18}}
\put(805,258){\makebox(0,0)[r]{$10.5$ cm}}
\put(404,503){\circle*{12}}
\put(506,506){\circle*{12}}
\put(609,555){\circle*{12}}
\put(881,258){\circle*{12}}
\thinlines \drawline[-50](199,494)(199,494)(207,494)(214,494)(222,494)(229,494)(237,494)(245,494)(252,494)(260,494)(267,494)(275,494)(283,494)(290,494)(298,494)(305,494)(313,494)(321,494)(328,494)(336,494)(344,494)(351,494)(359,494)(366,494)(374,494)(382,494)(389,494)(397,494)(404,494)(412,494)(420,494)(427,494)(435,494)(442,494)(450,494)(458,494)(465,494)(473,494)(480,494)(488,494)(496,494)(503,494)(511,494)(518,494)(526,494)(534,494)(541,494)(549,494)(556,494)(564,494)(572,494)
\thinlines \drawline[-50](572,494)(579,494)(587,494)(595,494)(602,494)(610,494)(617,494)(625,494)(633,494)(640,494)(648,494)(655,494)(663,494)(671,494)(678,494)(686,494)(693,494)(701,494)(709,494)(716,494)(724,494)(731,494)(739,494)(747,494)(754,494)(762,494)(769,494)(777,494)(785,494)(792,494)(800,494)(807,494)(815,494)(823,494)(830,494)(838,494)(846,494)(853,494)(861,494)(868,494)(876,494)(884,494)(891,494)(899,494)(906,494)(914,494)(922,494)(929,494)(937,494)(944,494)(952,494)
\end{picture}
  \end{center}
  \caption{}
   \label{fig: range}
\end{figure}

\begin{figure}[htb]
  \begin{center}
\setlength{\unitlength}{0.240900pt}
\begin{picture}(1500,900)(0,0)
\thicklines \path(199,179)(219,179)
\thicklines \path(952,179)(932,179)
\put(177,179){\makebox(0,0)[r]{0}}
\thicklines \path(199,292)(219,292)
\thicklines \path(952,292)(932,292)
\put(177,292){\makebox(0,0)[r]{}}
\thicklines \path(199,404)(219,404)
\thicklines \path(952,404)(932,404)
\put(177,404){\makebox(0,0)[r]{0.4}}
\thicklines \path(199,517)(219,517)
\thicklines \path(952,517)(932,517)
\put(177,517){\makebox(0,0)[r]{}}
\thicklines \path(199,630)(219,630)
\thicklines \path(952,630)(932,630)
\put(177,630){\makebox(0,0)[r]{0.8}}
\thicklines \path(199,742)(219,742)
\thicklines \path(952,742)(932,742)
\put(177,742){\makebox(0,0)[r]{}}
\thicklines \path(199,855)(219,855)
\thicklines \path(952,855)(932,855)
\put(177,855){\makebox(0,0)[r]{1.2}}
\thicklines \path(199,179)(199,199)
\thicklines \path(199,855)(199,835)
\put(199,134){\makebox(0,0){0}}
\thicklines \path(378,179)(378,199)
\thicklines \path(378,855)(378,835)
\put(378,134){\makebox(0,0){0.05}}
\thicklines \path(558,179)(558,199)
\thicklines \path(558,855)(558,835)
\put(558,134){\makebox(0,0){0.1}}
\thicklines \path(737,179)(737,199)
\thicklines \path(737,855)(737,835)
\put(737,134){\makebox(0,0){0.15}}
\thicklines \path(916,179)(916,199)
\thicklines \path(916,855)(916,835)
\put(916,134){\makebox(0,0){0.2}}
\thicklines \path(199,179)(952,179)(952,855)(199,855)(199,179)
\put(45,517){\makebox(0,0)[l]{\shortstack{$\zeta/d$}}}
\put(575,67){\makebox(0,0){$d/R$}}
\put(805,742){\makebox(0,0)[r]{$R=3.5$ cm}}
\thinlines \path(827,742)(935,742)
\thinlines \path(827,752)(827,732)
\thinlines \path(935,752)(935,732)
\thinlines \path(916,287)(916,332)
\thinlines \path(906,287)(926,287)
\thinlines \path(906,332)(926,332)
\thinlines \path(814,287)(814,319)
\thinlines \path(804,287)(824,287)
\thinlines \path(804,319)(824,319)
\thinlines \path(711,275)(711,309)
\thinlines \path(701,275)(721,275)
\thinlines \path(701,309)(721,309)
\thinlines \path(660,282)(660,307)
\thinlines \path(650,282)(670,282)
\thinlines \path(650,307)(670,307)
\thinlines \path(609,297)(609,331)
\thinlines \path(599,297)(619,297)
\thinlines \path(599,331)(619,331)
\thinlines \path(558,322)(558,348)
\thinlines \path(548,322)(568,322)
\thinlines \path(548,348)(568,348)
\thinlines \path(506,333)(506,378)
\thinlines \path(496,333)(516,333)
\thinlines \path(496,378)(516,378)
\thinlines \path(455,384)(455,411)
\thinlines \path(445,384)(465,384)
\thinlines \path(445,411)(465,411)
\thinlines \path(404,438)(404,472)
\thinlines \path(394,438)(414,438)
\thinlines \path(394,472)(414,472)
\thinlines \path(353,536)(353,573)
\thinlines \path(343,536)(363,536)
\thinlines \path(343,573)(363,573)
\thinlines \path(301,799)(301,844)
\thinlines \path(291,799)(311,799)
\thinlines \path(291,844)(311,844)
\put(916,309){\makebox(0,0){$\star$}}
\put(814,303){\makebox(0,0){$\star$}}
\put(711,292){\makebox(0,0){$\star$}}
\put(660,294){\makebox(0,0){$\star$}}
\put(609,314){\makebox(0,0){$\star$}}
\put(558,335){\makebox(0,0){$\star$}}
\put(506,356){\makebox(0,0){$\star$}}
\put(455,398){\makebox(0,0){$\star$}}
\put(404,455){\makebox(0,0){$\star$}}
\put(353,555){\makebox(0,0){$\star$}}
\put(301,821){\makebox(0,0){$\star$}}
\put(881,742){\makebox(0,0){$\star$}}
\put(805,697){\makebox(0,0)[r]{$7$ cm}}
\put(353,558){\circle{18}}
\put(404,471){\circle{18}}
\put(506,368){\circle{18}}
\put(609,328){\circle{18}}
\put(711,303){\circle{18}}
\put(814,310){\circle{18}}
\put(916,300){\circle{18}}
\put(881,697){\circle{18}}
\put(805,652){\makebox(0,0)[r]{$10.5$ cm}}
\put(404,463){\circle*{12}}
\put(506,370){\circle*{12}}
\put(609,343){\circle*{12}}
\put(881,652){\circle*{12}}
\put(805,607){\makebox(0,0)[r]{exp-fit}}
\thinlines \path(827,607)(935,607)
\thinlines \path(294,855)(298,828)(305,784)(313,743)(321,706)(328,672)(336,641)(344,612)(351,586)(359,562)(366,540)(374,520)(382,501)(389,485)(397,469)(404,455)(412,442)(420,430)(427,419)(435,409)(442,400)(450,391)(458,383)(465,376)(473,370)(480,364)(488,358)(496,353)(503,349)(511,345)(518,341)(526,337)(534,334)(541,331)(549,328)(556,326)(564,324)(572,322)(579,320)(587,318)(595,316)(602,315)(610,313)(617,312)(625,311)(633,310)(640,309)(648,308)(655,307)(663,307)(671,306)
\thinlines \path(671,306)(678,305)(686,305)(693,304)(701,304)(709,303)(716,303)(724,303)(731,302)(739,302)(747,302)(754,301)(762,301)(769,301)(777,301)(785,301)(792,300)(800,300)(807,300)(815,300)(823,300)(830,300)(838,300)(846,300)(853,300)(861,300)(868,299)(876,299)(884,299)(891,299)(899,299)(906,299)(914,299)(922,299)(929,299)(937,299)(944,299)(952,299)
\end{picture}
  \end{center}
  \caption{}
   \label{fig: range3}
\end{figure}

\fi

\end{document}